\documentclass{article}

\usepackage[utf8]{inputenc}
\usepackage[english]{babel}
\usepackage[top=3.5cm, bottom=3.5cm, right=3.5cm, left=3.5cm]{geometry}
\usepackage{hyperref}
\usepackage{graphicx}%
\usepackage{multirow}%
\usepackage{amsmath,amssymb,amsfonts}%
\usepackage{amsthm}%
\usepackage{mathrsfs}%
\usepackage[title]{appendix}%
\usepackage[x11names]{xcolor}%
\usepackage{textcomp}%
\usepackage{manyfoot}%
\usepackage{booktabs}%
\usepackage{algorithm}%
\usepackage{algorithmicx}%
\usepackage{algpseudocode}%
\usepackage{listings}%
\usepackage{tikz}%
\usepackage{textcomp}%
\usepackage{paralist}%

\usetikzlibrary{positioning} 
\usetikzlibrary{shapes} 
\usetikzlibrary{arrows,decorations.markings} 
\tikzstyle directed=[postaction={decorate,decoration={markings,mark=at position .65 with {\arrow[scale=1.2]{>}}}}]

\tikzstyle{every matrix}=[ampersand replacement=\&]
\tikzstyle{shorthandoff}=[]
\tikzstyle{shorthandon}=[]

\newtheorem{thm}{Theorem}
\newtheorem{prop}{Proposition}

\newtheorem{example}{Example}
\newcommand{\B}{\mathbb B}
\newcommand{\N}{\mathbb N}

\newcommand{\R}{\mathbb R}

\usepackage{listings}
\begin{document}

\title{Robustness of Boolean networks to update modes: an application to hereditary angioedema}
	
\author{Jacques Demongeot\,$^{1,}$\thanks{
		\href{mailto:jacques.demongeot@univ-grenoble-alpes.fr}{jacques.demongeot@univ-grenoble-alpes.fr}
	}~, Eric Goles\,$^{2,}$\thanks{
	  \href{mailto:eric.chacc@uai.cl}{eric.chacc@uai.cl}
	}~, Houssem ben Khalfallah\,$^{1,}$\thanks{
	  \href{mailto:houssem.ben-khalfallah@univ-grenoble-alpes.fr}{houssem.ben-khalfallah@univ-grenoble-alpes.fr}
	}~,\\
	Marco Montalva-Medel\,$^{2,}$\thanks{
	  \href{mailto:marco.montalva@uai.cl}{marco.montalva@uai.cl}
	}~ and Sylvain Sen{\'e}\,$^{3,4,}$\thanks{
		\href{mailto:sylvain.sene@lis-lab.fr}{sylvain.sene@lis-lab.fr}
	}\\[2mm]
	{\small $^1$~Universit{\'e} Grenoble Alpes, Avenue des Maquis du Graisivaudan, La Tronche, 
	  38700, France}\\
	{\small $^2$~Universidad Adolfo Ib{\'a}{\~n}ez, Av. Diagonal Las Torres 2640, 
	Pe{\~n}alol{\'e}n, 7910133, Santiago, Chile}\\
	{\small $^3$~Aix Marseille Univ, CNRS, LIS, Marseille, France}\\
	{\small $^4$~Universit{\'e} publique, Marseille, France}
}

\date{}

\maketitle

\begin{abstract}
Many familial diseases are caused by genetic accidents, which affect both the genome and its 
epigenetic environment, expressed as an interaction graph between the genes as that involved in one 
familial disease we shall study, the hereditary angioedema. 
The update of the gene states at the vertices of this graph (1 if a gene is activated, 0 if it is 
inhibited) can be done in multiple ways, well studied over the last two decades: Parallel, 
sequential, block-sequential, block-parallel, random, etc. 
We will study a particular graph, related to the familial disease proposed as an example, which has 
subgraphs which activate in an intricate manner (\emph{i.e.}, in an alternating block-parallel mode, 
with one core constantly updated and two complementary subsets of genes alternating their updating), 
of which we will study the structural aspects, robust or unstable, in relation to some classical 
periodic update modes.\\[2mm]
\end{abstract}


\section{Introduction}
\label{sec:intro}

The problem of updating states in a dynamical system is difficult to take into account in 
applications. 
In continuous systems defined by means of ordinary differential equations (ODEs), it is possible to 
take into account various forms of sequentiality, from massive parallelism, to random sequentiality, 
passing through specific schemes of activation of state variables taking into account the 
specificities of the simulated real system, for example a neural network~\cite{Mayo1989}. 
By default, the Runge-Kutta program for ODE simulation usually follows a parallel scheme, in which 
all states are updated based on the state values at the previous iteration. 
In discrete systems, for example in Boolean neural or genetic networks, the problem of state 
updating is crucial.\medskip

Since 45 years, numerous studies both theoretical and applied have shown that the asymptotic 
behaviors of the trajectories (called attractors) of discrete systems is largely dependent on how 
the states of the variables constituting these systems are 
updated~\cite{Goles1980,Aracena2004,Demongeot2020}. 
However, it is these asymptotic behaviors that are crucial in applications, particularly in 
epigenetic control problems. In these problems, a large number of elements interact (genes, 
proteins, RNAs, etc.) and, initially, their presence (concentration above a certain threshold) and 
their absence (concentration below this threshold) are symbolized by state 1 and state 0, 
respectively. 
This is referred to as a dynamical Boolean genetic control system. 
The study of attractors and their basins of attraction is essential in a Boolean dynamical system, 
as some correspond to normal physiological behavior and others to pathological behaviors that must 
be anticipated in order to prevent the onset of the corresponding pathologies. 
If the nature (constant, periodic or chaotic) of the attractors and the width of their basin depend 
on how the discrete system under study is updated, then any preventive or therapeutic approach will 
also depend on this.\medskip

Let us now consider a real genetic control system. 
Setting the variable corresponding to a given gene to state 1 depends on its transcription into 
messenger RNA. 
For this to happen, its sequence in the DNA of the cell nucleus where it will be expressed must be 
read and exported to the cytoplasm of that cell in the form of messenger RNA. 
This step is controlled by proteins called histones. 
If they are present in large numbers, many genes can be transcribed at the same time. 
The transcription mechanism can then be described as parallel. 
If histones are few in number or inhibited by transcription factors, genes will be transcribed in an 
order that depends on their ability to access the DNA of the genes in the nucleus. 
The corresponding dynamics are then referred to as block-sequential (parallel in a cotranscription 
block, with blocks being transcribed in a certain order sequentially). 
It is therefore easy to see that the concentration of the histone pool (sometimes called the 
chromatin clock) will have a considerable influence on the asymptotic expression of genes, 
especially since their messenger RNAs can still be inhibited, before being translated into proteins 
in the ribosomes of the cytoplasm, by small RNAs called microRNAs.\smallskip

If we consider that the histones of the chromatin clock are themselves proteins resulting from 
transcription and translation, it is easy to understand that their concentration may depend on 
transcription factors that may be RNAs or proteins whose dynamics they themselves control. 
This is referred to as state-dependent updating, since the clock depends on the state of the 
system's variables. 
We can therefore observe complex dynamics similar to those of non-autonomous differential systems 
with updating modes that can be described as entangled or intricate, in which each variable has its 
own clock.
The study of Boolean dynamics with intricate or state-dependent updates is thus crucial in the case 
of their application in epigenetic control systems.\medskip

In this paper, we present an interaction network involved in hereditary angioedema, which is a 
familial genetic disease which, as other, is due to epigenetic ``accidents'' caused on the genome of 
the first pathological carrier. 
Then, we study the dynamical behaviors of the underlying genetic regulatory network with a large 
variety of periodic update modes, by basing ourselves on Boolean automata network modeling 
and finite dynamical system theory. 
This leads us to show that our model is particularly robust to variations in synchronism, which 
allows us to infer that if changes are observed in terms of gene expression in hereditary 
angioedema, they are unlikely to stem from the influence of biological clocks but are more likely 
due to other phenomena, such as hormonal influences or exogenous factors.\medskip


After introducing the mathematical context of the Boolean network modeling in 
Section~\ref{sec:context}, Section~\ref{sec:results} is dedicated to the description of the 
interaction graph of the genetic regulatory network, from which a functional core subgraph is 
extracted and its block-sequential dynamical robustness is studied \emph{per se}. 
This section ends with a discussion around a specific intricate update mode, chosen from the 
structure of the network itself, which captures as a unique attractor one the most frequent limit 
cycles obtained by the diversity of block-sequential update modes. 
Section~\ref{sec:conclusion} is devoted to the conclusion.

\section{Mathematical context}\label{sec:context}

\subsection{Definitions}\label{sec:context:definitions}

\subsubsection{Basic notations}

Let $n \geq 1 \in \N$ be a positive integer, $\B = \{0,1\}$ the set of Boolean numbers, $x$ a vector 
of $\B^n$ with $x_i$ its $i$-th component, $x_L$ the projection of $x$ onto an element of $\B^{|L|}$ 
for some subset $L$ of $N = \{1, \dots, n\}$. 
In the following, for $a, b \in \B, \neg a, a \land b, a \lor b$ denote respectively the 
opposite of $a$, the infimum $\inf(a,b)$ and the supremum $\sup(a,b)$  between $a$ and $b$.

\subsubsection{Boolean networks}

A Boolean automata network (BAN) of size $n$, abbreviated by BAN$_n$, is a model of discrete 
dynamical systems composed of a set of $n$ automata, the $i$-th ($i \in N$) holding a state $x_i 
\in \B$. 
This model was introduced in~\cite{Kauffman1969,Thomas1973} on the basis of the seminal work 
of McCulloch and Pitts~\cite{McCulloch1943} on formal neural networks. 
A component of a BAN is called an automaton, because it is an element which computes its own state 
based on that of its neighbors over discrete time. 
A configuration of a BAN$_n$ is an element of $\B^n$, which can also be encoded as a word on the 
alphabet $\B$ of length $n$ such that, given for instance $n = 5$, $x = (0, 1, 0, 1, 1) \equiv 
01011$. 
A BAN$_n$ is defined by a global function $f: \B^n \to \B^n$, decomposed into $n$ local functions 
$f_i: \B^n \to \B$, $i \in N$, where $f_i$ is the $i$-th component of $f$.\medskip

Information provided by the local functions can be summarized by a directed and signed graph, called 
the \emph{interaction graph} of BAN$_n$ $f$, $G_f = (V,E)$ with its vertices $V = N$ and edges 
$E \subseteq N \times \{+, -\} \times N$.
Whenever $(i, +, j) \in E$, we say that $i$ is an \emph{activator} of $j$;
whenever $(i, -, j) \in E$, $i$ is an \emph{inhibitor} of $j$.
Notice that automaton $i$ can be both an activator and an inhibitor of $j$ when the local function 
$f_j$ is locally non-monotone on $i$.
The interaction graph of $f$ has a positive (resp. negative) edge from $i$ to $j$ if and only if one 
can assign a state to each automaton other than $i$ so that the local function of $j$ becomes equal 
(resp. different) to the state of $i$.\medskip

Given a BAN, to let it evolve over discrete time, one must define when the automata update their 
state using their local function, which can be done in multiple ways as introduced 
by Robert~\cite{Robert1986} and recently reviewed in~\cite{Pauleve2022}.\medskip

\subsubsection{Block-sequential update modes}

When we have a BAN$_n$, the informal idea behind a block-sequential update mode of it consists in 
separating the set of automata in disjoints blocks into a sequence and iterating sequentially the 
blocks. 
When a block is iterated, each of its automata are updated in parallel (\emph{i.e.} 
synchronously).\medskip

A sequence $(W_i)_{i \in M = \{1, \dots, m\}, m \leq n}$ of subsets $W_i$ of $N$ is an 
\emph{ordered partition} if and only if:
\begin{equation*}
  \forall i \in M, W_i \neq \emptyset;\; \cup_{i \in M} W_i = N;\; \text{and } \forall i, j \in M, 
  i \neq j \implies W_i \cap W_j = \emptyset\text{.}
\end{equation*}

A sequence $\mu = (W_i)_{i \in M}$ is a block-sequential update mode if and only if $\mu$ is an 
ordered partition of $N$, and the $W_i$ are called 
\emph{blocks}~\cite{Robert1986,Demongeot2008,Aracena2009}. 
The set of block-sequential update modes of a network of size $n$ is denoted by $\text{BS}_n$. 
The update of $f$ under $\mu \in \text{BS}_n$ is given by $f_\mu: \B^n \to \B^n$ as follows:
\begin{equation*}
  f_\mu(x) = f_{W_m} \circ \dots \circ f_{W_2} \circ f_{W_1}(x)\text{,}
\end{equation*}
where for all $i \in M$, for all $k \in N$, $f_{W_i}(x)_k = f_k(x)$, if $k \in W_i$, and 
$f_{W_i}(x)_k = x_k$ otherwise.

\begin{figure}[t!]
  \begin{center}
    \begin{minipage}{.3\textwidth}
			\centerline{$f: B^3 \to B^3$}\medskip
			
			\centerline{
			  $f(x) = \left\{\begin{array}{l}
			    f_1(x) = x_3\\
			    f_2(x) = x_1\\
			    f_3(x) = x_2\\
			  \end{array}\right.$
			}
		\end{minipage}
		\hspace*{20mm}
		\begin{minipage}{.3\textwidth}
			\centerline{\scalebox{1}{\begin{tikzpicture}[>=latex,auto]
				\tikzstyle{node} = [draw, circle, thick]
				\node[node](n1) at (0,0) {$1$};
				\node[node](n2) at (1,-1.5) {$2$};
				\node[node](n3) at (-1,-1.5) {$3$};
				\draw[Green4, thick, ->] (n1) edge [bend left=20] node {$+$} (n2);
				\draw[Green4, thick, ->] (n2) edge [bend left=20] node {$+$} (n3);
				\draw[Green4, thick, ->] (n3) edge [bend left=20] node {$+$} (n1);
			\end{tikzpicture}}}
		\end{minipage}\medskip
      
		\begin{minipage}{.3\textwidth}
			\centerline{(a)}
		\end{minipage}
		\hspace*{20mm}
		\begin{minipage}{.3\textwidth}
			\centerline{(b)}
		\end{minipage}
  \end{center}
  \caption{Boolean automata network $f$ of Example~\ref{ex:3pos-cycle} on 
    page~\pageref{ex:3pos-cycle}; 
    (a) its definition by means of functions; 
    (b) its associated interaction graph.}
  \label{fig:3pos-cycle}
\end{figure}
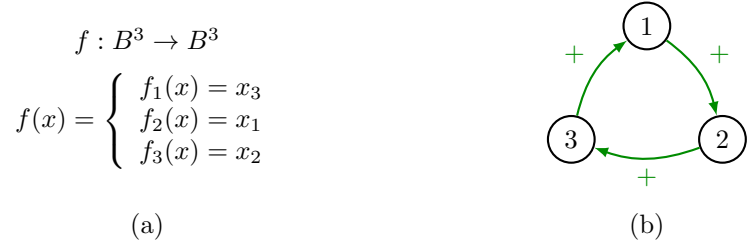

\subsubsection{Block-parallel update modes}

A block-parallel update mode is based on the dual principle ruling a block-sequential update mode, 
as explained in~\cite{Demongeot2020}: 
The automata in a same block are updated sequentially while the blocks are iterated in 
parallel.\medskip

Instead of being defined as a sequence of unordered blocks, a block-parallel update mode is thus 
defined as a set of ordered blocks. 
The set $\{S_j\}_{j \in K = \{1, \dots, k\}, k \leq n}$, where $S_j = (i_{j,1}, \dots, i_{j,n_j})$ 
is a sequence of $n_j > 0$ elements of $N$ for all $j \in K$, is a \emph{partitioned order} if and 
only if:
\begin{equation*}
  \forall j \in K, S_j \neq \emptyset;\; \cup_{j \in K} S_j = N;\; \text{and } \forall i, j \in K, 
  i \neq j \implies S_i \cap S_j = \emptyset\text{.}
\end{equation*}

A set $\mu = \{S_j\}_{j \in K}$ is a block-parallel mode if and only if $\mu$ is a partitioned 
order, and the $S_j$ are called \emph{o-blocks} (`o' for 
ordered)~\cite{Donoso2025,Perrot2024a,Perrot2024b}. 
The set of block-parallel update modes of a network of size $n$ is denoted by $\text{BP}_n$. 
If we denote by $p$ the lowest common multiple of $(n_1, \dots, n_k)$, the update of $f$ under 
$\mu \in \text{BP}_n$ is given by $f_\mu: \B^n \to \B^n$ as follows:
\begin{equation*}
  f_\mu(x) = f_{W_p} \circ \dots \circ f_{W_2} \circ f_{W_1}(x)\text{,}
\end{equation*}
where for all $j \in K$, we define $W_j = \{i_{j \mod n_k, k \in K}\}$.

\subsubsection{Intricate update modes}

Given an update mode written by means of a sequence of sets of updates, if the condition above 
``$W_i \cap W_j = \emptyset$'' is no more available, then the update mode is called 
\emph{intricate}, because an automaton can belong to several update sets and can thus be updated 
several times over a period. 
As a consequence, block-sequential update modes are not intricate by definition whereas 
block-parallel ones can be. 
The speed at which the state of an automaton evolve over time is indeed dependent on the number of 
update sets $W_i$ to which it belongs. 
Such intricate dynamics can cause attractors of the Boolean automata network to appear or disappear, 
or change their nature (\emph{e.g.} for limit cycles to stationary states). 
In applications, for example in genetics, we can think of precise molecular mechanisms linked to the 
temporality of clocks such as the chromatin clock, depending on circadian rhythms~\cite{Zhang2020}. 
The chromatin organization plays a crucial role in genetic regulation by controlling the 
accessibility of DNA to transcription machinery, related to the biosynthesis of proteins such as 
histones or polymerases. 
We can reasonably suppose that such a nuclear control of the update of gene states may 
lead to an intricate mode.

\subsubsection{Threshold Boolean automata network}

Threshold Boolean automata networks (TBANs) have been introduced in~\cite{McCulloch1943} and have 
been widely studied since decades in the context of computer science and 
mathematics~\cite{Kleene1951,Goles1980} as well as that of artificial 
intelligence~\cite{Hopfield1982}.
A TBAN of size $n$ is abbreviated by TBAN$_n$ and defined by a threshold function 
$f: \B^n \to \B^n$, decomposed into $n$ local functions $f_i: \B^n \to \B^n$, $i \in N$, where 
$f_i$ is the $i$-th component of $f$ defined by: 
\begin{equation*}
  \forall x \in \B^n, f_i(x) = H(\sum_{j=1}^n w_{ij} x_j - \theta_i)\text{,}
\end{equation*}
where $H$ is the Heaviside function defined as:
\begin{equation*}
  H(y) = \begin{cases}
    1 & \text{if } y \geq 0\\
    0 & \text{otherwise}
  \end{cases} \text{,}
\end{equation*} 
and $w_{ij} \in \R$ is the interaction weight which represents the influence exerted by automaton 
$j$ on automaton $i$ in the TBAN$_n$. 
If $w_{ij} > 0$ (resp. $w_{ij} < 0$), the influence is called an \emph{activation} (resp. an 
\emph{inhibition}). 
If $w_{ij} = 0$, automaton $i$ does not depend on automaton $j$.\medskip

The interaction graph $G_f$ of a TBAN$_n$ $f$ is similar to that of a BAN$_n$. 
However, each interaction is characterized by the weight $w_{ij}$ which represents the influence 
that a source automaton $j$ exerts on an target automaton $i$ of the graph $G$.
If the automata represent genes, this influence of gene $j$ on gene $i$ can be exerted by the 
messenger RNA or the protein that gene $j$ expresses. 

\subsubsection{Attractors}

In life sciences (namely biology and medicine) the primitive notion of an attractor was proposed 
by Buffon~\cite{Buffon1749} under the name of ``inner mold'' (``moule int{\'e}rieur''). 
Then, many other authors refined his primitive intuition: Birkhoff was the first to introduce the 
notion of a center of attraction of a dynamical system in~\cite{Birkhoff1927}. 
He was followed by numerous authors~\cite{Smale1967,Bhatia1967,Thom1972,Williams1974}.\medskip

Then, the multiplication of numerical experiments on strange attractors has shown that the previous 
definitions were too restrictive. 
Using the notion of Bowen's shadow trajectory described in~\cite{Bowen1975}, numerous works 
refined the first definitions~\cite{Conley1978,Ruelle1981,Sinai1981,Guckenheimer1983,Garrido1983}. 
Then, Cosnard and Demongeot proposed a more general approach to the concept of an 
attractor~\cite{Cosnard1985}, sought to broaden all these definitions by adding the following 
essential characteristics:
\begin{enumerate}
\item The basin of the attractor contains a strict neighborhood of it.
\item The attractor verifies a minimality condition.
\item It is generally assumed to be closed and invariant by the flow of the dynamics.
\end{enumerate}

These characteristics can be translated mathematically into the following framework. 
Let consider a temporal set $T \subseteq \R_+$ and a state space $E \subseteq \R^n$ provided with a 
dynamical flow $\varphi$ on $E \times T$ and a distance $d$. 
Then, sequence $(\varphi(a,t))_{t \in T}$ denotes the trajectory starting at state $a$ in $E$ and 
being at state $\varphi(a,t)$ at time $t$. 
We define reciprocal operators $L$ and $B$ on the set of subsets of $E$. 
$L(a)$ denotes the limit set of the trajectory starting in $a$, made of the accumulation points of 
the trajectory $(\varphi(a,t))_{t \in T}$, when $t$ tends to infinity:
\begin{multline}
  L(a) = \{y \in E \mid \forall \epsilon > 0, \forall t \in T, \exists s(\epsilon, t) \in T\\
  \text{ such that } s(\epsilon, t) > t \text{ and } d(\varphi(a, s(\epsilon, t)), y) \leq 
  \epsilon\}\text{,}
\end{multline}
where $s(\epsilon, t)$ is a time after $t$ at which the trajectory starting in $a$ is in a ball of 
radius $\epsilon$ centered on $y$.\medskip

Considering a subset $A$ of $E$, $L(A)$ is the union of all limit sets $L(a)$, for $a$ belonging to 
$A$ such that:
\begin{equation}
  L(A) = \cup_{a\in A} L(a)\text{.}
\end{equation}

\begin{figure}[t!]
	\begin{center}
    \begin{minipage}{.3\textwidth}
      \centerline{\scalebox{1}{\begin{tikzpicture}[>=to,auto]
        	\tikzstyle{type} = []
        \tikzstyle{conf} = [rectangle, draw]
        \tikzstyle{pf} = [rectangle, draw, fill=black!10]
        \tikzstyle{lc} = [rectangle, draw, fill=black!70]
        \node[type](titre) at (0,3) {(a)};
        \node[pf](n000) at (0,0) {$000$};
        \node[lc](n001) at (2,0) {\textcolor{white}{$001$}};
        \node[lc](n010) at (0,2) {\textcolor{white}{$010$}};
        \node[lc](n011) at (2,2) {\textcolor{white}{$011$}};
        \node[lc](n100) at (1,1) {\textcolor{white}{$100$}};
        \node[lc](n101) at (3,1) {\textcolor{white}{$101$}};
        \node[lc](n110) at (1,3) {\textcolor{white}{$110$}};
        \node[pf](n111) at (3,3) {$111$};
        \draw[->] (n000) edge [loop above, distance=4mm] (n000);
        \draw[->] (n001) edge (n100);
        \draw[->] (n010) edge [bend right=40] (n001);
        \draw[->] (n011) edge (n101);
        \draw[->] (n100) edge (n010);
        \draw[->] (n101) edge [bend right=40] (n110);
        \draw[->] (n110) edge (n011);
        \draw[->] (n111) edge [loop below, distance=4mm] (n111);
      \end{tikzpicture}}}
    \end{minipage}
    \hspace*{20mm}
    \begin{minipage}{.3\textwidth}
      \centerline{\scalebox{1}{\begin{tikzpicture}[>=to,auto]
        	\tikzstyle{type} = []
        \tikzstyle{conf} = [rectangle, draw]
        \tikzstyle{pf} = [rectangle, draw, fill=black!10]
        \tikzstyle{lc} = [rectangle, draw, fill=black!70]
        \node[type](titre) at (0,3) {(b)};
        \node[pf](n000) at (0,0) {$000$};
        \node[conf](n001) at (2,0) {$001$};
        \node[conf](n010) at (0,2) {$010$};
        \node[lc](n011) at (2,2) {\textcolor{white}{$011$}};
        \node[lc](n100) at (1,1) {\textcolor{white}{$100$}};
        \node[conf](n101) at (3,1) {$101$};
        \node[conf](n110) at (1,3) {$110$};
        \node[pf](n111) at (3,3) {$111$};
        \draw[->] (n000) edge [loop right, distance=4mm] (n000);
        \draw[->] (n001) edge (n100);
        \draw[->] (n010) edge (n000);
        \draw[->] (n011) edge [bend right=20] (n100);
        \draw[->] (n100) edge [bend right=20] (n011);
        \draw[->] (n101) edge (n111);
        \draw[->] (n110) edge (n011);
        \draw[->] (n111) edge [loop left, distance=4mm] (n111);
      \end{tikzpicture}}}
    \end{minipage}
  \end{center}\bigskip\bigskip
  
  \begin{center}
    \begin{minipage}{.3\textwidth}
      \centerline{\scalebox{1}{\begin{tikzpicture}[>=to,auto]
        	\tikzstyle{type} = []
        \tikzstyle{conf} = [rectangle, draw]
        \tikzstyle{pf} = [rectangle, draw, fill=black!10]
        \tikzstyle{lc} = [rectangle, draw, fill=black!70]
        \node[type](titre) at (0,3) {(c)};
        \node[pf](n000) at (0,0) {$000$};
        \node[conf](n001) at (2,0) {$001$};
        \node[lc](n010) at (0,2) {\textcolor{white}{$010$}};
        \node[conf](n011) at (2,2) {$011$};
        \node[conf](n100) at (1,1) {$100$};
        \node[lc](n101) at (3,1) {\textcolor{white}{$101$}};
        \node[conf](n110) at (1,3) {$110$};
        \node[pf](n111) at (3,3) {$111$};
        \draw[->] (n000) edge [loop above, distance=4mm] (n000);
        \draw[->] (n001) edge [bend left=40] (n010);
        \draw[->] (n010) edge [bend right=10] (n101);
        \draw[->] (n011) edge (n111);
        \draw[->] (n100) edge (n000);
        \draw[->] (n101) edge [bend right=10] (n010);
        \draw[->] (n110) edge [bend left=40] (n101);
        \draw[->] (n111) edge [loop below, distance=4mm] (n111);
      \end{tikzpicture}}}
    \end{minipage}
    \hspace*{20mm}
    \begin{minipage}{.3\textwidth}
      \centerline{\scalebox{1}{\begin{tikzpicture}[>=to,auto]
        	\tikzstyle{type} = []
        \tikzstyle{conf} = [rectangle, draw]
        \tikzstyle{pf} = [rectangle, draw, fill=black!10]
        \tikzstyle{lc} = [rectangle, draw, fill=black!70]
        \node[type](titre) at (0,3) {(d)};
        \node[pf](n000) at (0,0) {$000$};
        \node[conf](n001) at (2,0) {$001$};
        \node[pf](n010) at (0,2) {$010$};
        \node[conf](n011) at (2,2) {$011$};
        \node[conf](n100) at (1,1) {$100$};
        \node[pf](n101) at (3,1) {$101$};
        \node[conf](n110) at (1,3) {$110$};
        \node[pf](n111) at (3,3) {$111$};
        \draw[->] (n000) edge [loop above, distance=4mm] (n000);
        \draw[->] (n001) edge (n000);
        \draw[->] (n010) edge [loop above, distance=4mm] (n010);
        \draw[->] (n011) edge (n010);
        \draw[->] (n100) edge (n101);
        \draw[->] (n101) edge [loop below, distance=4mm] (n101);
        \draw[->] (n110) edge (n111);
        \draw[->] (n111) edge [loop below, distance=4mm] (n111);
      \end{tikzpicture}}}
    \end{minipage}
  \end{center}
  \caption{Different periodic dynamics of the Boolean automata network of 
    Example~\ref{ex:3pos-cycle} on page~\pageref{ex:3pos-cycle} defined in 
    Figure~\ref{fig:3pos-cycle}; 
    (a) its parallel dynamics according to $\mu_\textsc{par} = (\{1, 2, 3\})$; 
    (b) its block-sequential dynamics according to $\mu_\textsc{bs} = (\{1, 2\}, \{3\})$;
    (c) its block-parallel dynamics according to $\mu_\textsc{bp} = \{(1), (3, 2)\} \equiv 
      (\{1, 3\}, \{1, 2\})$;
    (d) its intricate dynamics according to $\mu_\textsc{in} = (\{2, 3\}, \{1, 3\}, \{1,2\})$.}
  \label{fig:dyn-3pos-cycle}
\end{figure}
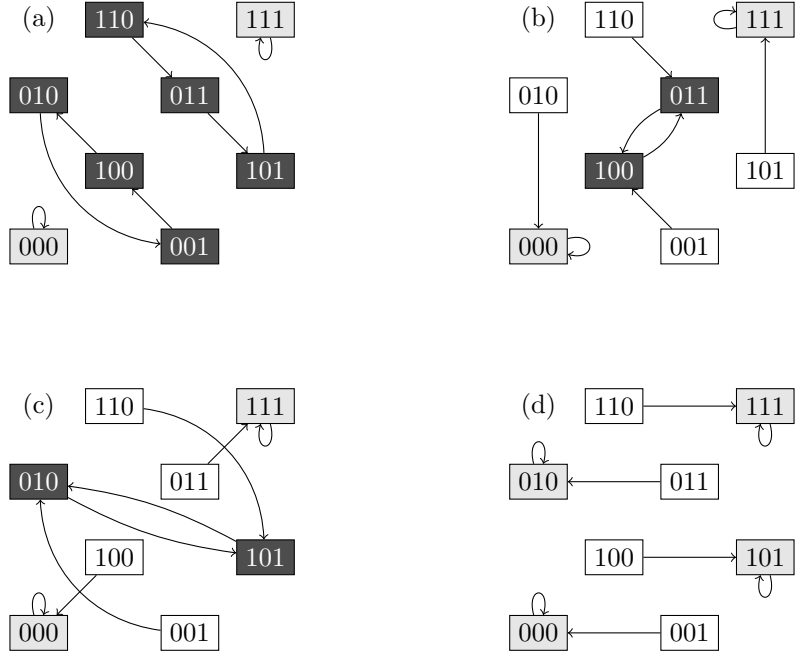

Conversely, $B(A)$ is the set of all initial conditions $a$ outside $A$, whose limit set $L(a)$ is 
included in $A$ such that:
\begin{equation} 
  B(A) = \{y \in E \setminus A \mid L(y) \subseteq A)\}\text{,}
\end{equation} 
and is called the \emph{attraction basin} of $A$. 
Then, $A$ is an attractor if it verifies the three following conditions, where the notion of a 
shadow connection is defined in~\cite{Bowen1975}:
\begin{enumerate}
\item $A = L(B(A))$.
\item There is no set $A'$ containing strictly $A$ and shadow-connected to $A$.
\item There is no set $A''$ strictly contained in $A$ verifying 1. and 2.
\end{enumerate}

Using these definitions under the transition function $f$, configuration $x(t)$ of a BAN at time 
$t$ of its dynamics can evolve in $\B^n$ toward two possible asymptotic behaviors, \emph{i.e.}, two 
possible attractors: 
A fixed configuration (also called a fixed point or a stationary state) or a cycle of configurations 
(also called a limit cycle).\medskip
 
One of the challenges in a genetic regulatory network is to determine the nature of its attractors 
(fixed points or limit cycles), which can have a great influence on its physiological or 
pathological character. 
In the applications, it is crucial to identify all the possible attractors and their attraction 
basins (at least their size).
Note that, abusing language in the sequel, given a limit cycle (which can be a fixed point) $C$, for 
the sake of simplicity in the presentation of results with no loss of coherence, we consider that 
$C$ belongs to its attraction basin $B(A)$.

\begin{example}\normalsize\normalfont\label{ex:3pos-cycle}
  Let us consider a BAN$_3$, whose local functions are of arity $1$ and interactions between 
  automata can be depicted by a cycle, as illustrated in Figure~\ref{fig:3pos-cycle}.
  This BAN admits $(0, 0, 0)$ and $(1, 1, 1)$ as stationary states whatever the update mode because 
  these two configurations are fixed points of the global function $f$.
  Under block-sequential update modes, it admits also one or two limit cycles depending on the 
  chosen mode (\emph{e.g.}, two in the case of one block, called the parallel update mode). 
  The intricate update mode defined by $\mu = (\{2, 3\}, \{1, 3\}, \{1, 2\})$, which belongs to 
  neither BS$_3$ nor BP$_3$, has four stationary states~\cite{Demongeot2020}, which implies the 
  creation of two new stationary states with respect to the block-sequential update mode
  $(\{1, 2\}, \{3\})$ or the block-parallel update mode $\{(1), (3, 2)\}$ for instance, which have 
  both two stationary states and one limit cycle. 
  Figure~\ref{fig:dyn-3pos-cycle} illustrates these different dynamics thanks to directed graphs 
  (commonly called \emph{transition graphs} in the domain) in which vertices are configurations and 
  an edge from configuration $x$ to configuration $y$ denotes that $y$ is the image of $x$ by the 
  flow of the underlying dynamical systems. In such transition graphs, fixed points are 
  configurations in a gray rectangle and configurations which belong to limit cycles are in a black 
  rectangle. 
  If we notice that generating fixed points which are not fixed points of the global functions $f$ 
  with a block-parallel update mode of a positive cycle needs at least $5$ 
  automata~\cite{Perrot2026}, then the intricate mode $\mu$ can be considered as providing``new'' 
  attractors ``at least cost''. 
\end{example}

%
%
%

\subsubsection{Robustness}

An updating robustness study of network dynamics consists in considering all the possible update 
modes (of a certain family, because the set of all possible deterministic update modes is 
infinite and uncountable) and showing which changes of states can occur when the update mode 
changes. 
The network can be robust for five types of perturbations -- change of initial conditions, parameter 
values, interaction graph, transition function or update mode -- and three types of stability:
\begin{enumerate}
\item The trajectorial (or Lyapunov) stability, which corresponds to the existence of a respected 
  distance threshold between the ancient trajectory and the new after perturbation.
\item The asymptotic stability, which corresponds to the conservation of an attractor after a 
  perturbation of initial conditions in its attraction basin.
\item The structural stability, which corresponds to the conservation of the number and the nature 
  of the attractors and their attraction basins, even if the transient part of trajectories changes 
  in response to structural perturbations (change of interaction graph, transition function or 
  update mode).
\end{enumerate}
In the following, we examine in the framework of a genetic application, the control of a 
familial genetic disease, the hereditary angioedema, the existence or absence of structural 
stability associated to the last of the five types of perturbation described above, the change of 
update mode, which is crucial in biology insofar as such an updating rule exists, because the exact 
functioning of the clock controlling the gene expression remains largely unknown.

\subsection{Network dynamics}\label{sec:context:dynamics}

Let consider the BAN$_n$ $f$ and $(l_1, \dots, l_k) \in \mathbb{N}^k$ for a given $k \in \{1, \dots, 2^n\}$. 
We say that:
\begin{itemize}
\item[--] A dynamics on $f$ is \emph{of type LC-$(l_1, \dots, l_k)$} if it has only $k$ limit cycles 
  of lengths $l_1$, $\dots$, $l_k$ respectively. (Notice that $l_i = 1$ if and only if the $i$-th 
  limit cycle is actually a fixed point).
\item[--] A set of configurations $I \subseteq \B^n$ is an \emph{intersector} of $A$, if every 
  dynamics of $f$ has at least one attractor whose all configurations belong to $I$. 
  A \emph{dominant set} $D \subseteq \B^n$ is a minimal intersector set of $A$. 
  An element $x \in D$ is called a \emph{dominant configuration} of $f$. 
  Similarly, an attractor (fixed point or limit cycle) whose configurations all belongs to $D$ is 
  a \emph{dominant attractor} of $f$.
\end{itemize}\medskip

In this work we analyze all the dynamics of a network, generated by all the block-sequential 
update schedules (in particular, that of the fully parallel update mode) which represent a big 
family of update modes that grows exponentially with the network size~\cite{Demongeot2008}.
Indeed, the number $|\text{BS}_n|$ of block-sequential update modes associated to a digraph of $n$ 
vertices is given by:
\begin{equation}
  |\text{BS}_n| = \sum_{k = 0}^{n-1} \binom{n}{k} |\text{BS}_k|\text{,}
\end{equation}
where $|\text{BS}_0| = 1$. 
If we add the fact that the dynamics produced by each of these update modes consists in $2^n$ 
configurations, then the amount of computations required for an exhaustive analysis of all these 
dynamics (which we perform in this work) is even greater. 
In practice, this analysis can only be conducted for small values of $n$. 
Next, we explain the main aspects of the update digraph theory which significantly reduces 
these computational costs.\medskip

Let us now summarize the main concepts and results developed 
in~\cite{Aracena2009,Aracena2011,Aracena2013b,Aracena2013a} which allow for the grouping of 
update modes generating exactly the same dynamics. 
Let us consider a digraph $G = (V, E)$, where $E$ is a set of directed edges between the vertices 
of $V = N$. 
If $\mu = (W_i)_{i \in M = \{1, \dots, m\}, m \leq n}$ is a block-sequential update mode, it can 
also be seen as a function $s: N \to M$. 
For instance, given a BAN$_4$, $\mu = (\{2\}, \{1,4\}, \{3\})$ can be seen as the function 
$s: \{1, 2, 3, 4\} \to \{1, 2, 3\}$ such that $s(1) = s(4) = 2$, $s(2) = 1$ and $s(3) = 3$. 
For the sake of generality, consider a BAN$_n$ and a block-sequential update mode $s$ on it. 
The \emph{label function} $\text{lab}_{s} : N \to \{<, \geq\}$ is defined by:
\begin{equation}
  \forall (i, j) \in E, \text{lab}_{s}(i, j) = 
    \begin{cases}
      < & \text{if } s(i) < s(j)\\
      \geq & \text{otherwise}
    \end{cases}\text{.}
\end{equation}

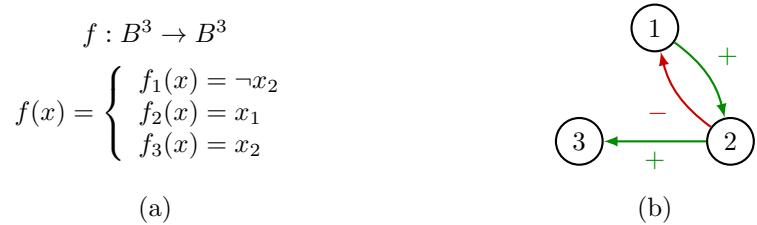
\begin{figure}[t!]
  \begin{center}
    \begin{minipage}{.3\textwidth}
      \centerline{$f: B^3 \to B^3$}\medskip
			
      \centerline{
			  $f(x) = \left\{\begin{array}{l}
			    f_1(x) = \neg x_2\\
			    f_2(x) = x_1\\
			    f_3(x) = x_2\\
			  \end{array}\right.$
			}
		\end{minipage}
		\hspace*{20mm}
    \begin{minipage}{.3\textwidth}
		  \centerline{\scalebox{1}{\begin{tikzpicture}[>=latex,auto]
        \tikzstyle{node} = [draw, circle, thick]
        \node[node](n1) at (0,0) {$1$};
        \node[node](n2) at (1,-1.5) {$2$};
        \node[node](n3) at (-1,-1.5) {$3$};
        \draw[Green4, thick, ->] (n1) edge [bend left=20] node {$+$} (n2);
        \draw[Red3, thick, ->] (n2) edge [bend left=20] node {$-$} (n1);
        \draw[Green4, thick, ->] (n2) edge node {$+$} (n3);
		  \end{tikzpicture}}}
		\end{minipage}\medskip
		
		\begin{minipage}{.3\textwidth}
		  \centerline{(a)}
		\end{minipage}
		\hspace*{20mm}
		\begin{minipage}{.3\textwidth}
		  \centerline{(b)}
		\end{minipage}
  \end{center}
  \caption{(a) Boolean automata network $f$ of Example~\ref{ex:UD}; (b) its associated interaction 
    graph $G = (V,E)$, where the set of vertices is $V = \{1, 2, 3\}$ and the set of edges is 
    $E = \{(1,2), (2,1), (2,3)\}$.}
  \label{fig:3digraph}
\end{figure}

By labeling every arc $(i, j)$ of $E$ with $<$ or $\geq$, we get a labeled digraph 
$(G, \text{lab})$. 
When $\text{lab} = \text{lab}_s$, for some update mode $s$, then the labeled digraph 
$(G, \text{lab}_s)$ is an \emph{update digraph}. 
Note that every update digraph is a labeled digraph, but the converse is not always true. 
In~\cite{Aracena2011}, the authors proved the following characterization for update digraphs.

\begin{thm}\normalsize\label{thm:arac2011}
  A labeled digraph is an update digraph if and only if changing the direction of only its arcs 
  with $<$-labels results in a new digraph (possibly a multidigraph) that does not contain any 
  cycle with a $<$-edge.
\end{thm}

Additionally, the following result was proven in~\cite{Aracena2009}, showing that update digraphs 
can be of great help to study the robustness of a Boolean automata network against changes of update 
mode.

\begin{thm}\normalsize\label{thm:arac2009}
  Let $f$ be a BAN, with $G_f = (V, E)$ its associated interaction graph. 
  Let $s_1$ and $s_2$ be two distinct block-sequential update modes defined on $V$, and 
  $(G_f, \text{lab}_{s_1})$ 
  and $(G_f, \text{lab}_{s_2})$ their related update digraphs. 
  The following result holds: If $(G_f, \text{lab}_{s_1}) = (G_f, \text{lab}_{s_2})$, then the 
  dynamics of $f$ under $s_1$ is exactly the same as that of $f$ under $s_2$.
\end{thm}

Using Theorem~\ref{thm:arac2009}, given a BAN$_n$ $f$ together with $G_f = (V, E)$, the following 
equivalence classes of block-sequential update modes that produce the same update digraph, and 
consequently the same dynamics for $f$, can be defined as:
\begin{equation}
  \forall s \in \text{BS}_n,\, [s]_{G_f} = 
  \{s' \mid (G_f, \text{lab}_{s'}) = (G_f, \text{lab}_s)\}\text{.}
\end{equation}
In this way, a bijection is established between the update digraphs and these equivalence classes, 
which partition the set of the $|\text{BS}_n|$ update modes, and consequently, allow for the study 
of all the distinct dynamics of a network while considering only a subset of the modes composed of 
the representative update modes of each class.\medskip

\begin{figure}[t!]
  \begin{center}
    \begin{minipage}{.3\textwidth}
      \centerline{\scalebox{1}{\begin{tikzpicture}[>=latex,auto]
        \tikzstyle{type} = []
        \tikzstyle{node} = [draw, circle, thick]
        \node[type](titre) at (-1,0) {(a)};
        \node[node](n1) at (0,0) {$1$};
        \node[node](n2) at (1,-1.5) {$2$};
        \node[node](n3) at (-1,-1.5) {$3$};
        \draw[thick, ->] (n1) edge [bend left=20] node {$\geq$} (n2);
        \draw[thick, ->] (n2) edge [bend left=20] node {$\geq$} (n1);
        \draw[thick, ->] (n2) edge node {$\geq$} (n3);
		  \end{tikzpicture}}}
    \end{minipage}
    \hfill
    \begin{minipage}{.3\textwidth}
      \centerline{\scalebox{1}{\begin{tikzpicture}[>=latex,auto]
        \tikzstyle{type} = []
        \tikzstyle{node} = [draw, circle, thick]
        \node[type](titre) at (-1,0) {(b)};
        \node[node](n1) at (0,0) {$1$};
        \node[node](n2) at (1,-1.5) {$2$};
        \node[node](n3) at (-1,-1.5) {$3$};
        \draw[thick, ->] (n1) edge [bend left=20] node {$\geq$} (n2);
        \draw[thick, ->] (n2) edge [bend left=20] node {$\geq$} (n1);
        \draw[LightSalmon4, thick, ->] (n2) edge node {$<$} (n3);
      \end{tikzpicture}}}
    \end{minipage}
    \hfill
    \begin{minipage}{.3\textwidth}
      \centerline{\scalebox{1}{\begin{tikzpicture}[>=latex,auto]
        	\tikzstyle{type} = []
        \tikzstyle{node} = [draw, circle, thick]
        \node[type](titre) at (-1,0) {(c)};
        \node[node](n1) at (0,0) {$1$};
        \node[node](n2) at (1,-1.5) {$2$};
        \node[node](n3) at (-1,-1.5) {$3$};
        \draw[LightSalmon4, thick, ->] (n1) edge [bend left=20] node {$<$} (n2);
        \draw[thick, ->] (n2) edge [bend left=20] node {$\geq$} (n1);
        \draw[thick, ->] (n2) edge node {$\geq$} (n3);
      \end{tikzpicture}}}
    \end{minipage}
  \end{center}\bigskip\bigskip

  \begin{center}
    \begin{minipage}{.3\textwidth}
      \centerline{\scalebox{1}{\begin{tikzpicture}[>=latex,auto]
        \tikzstyle{type} = []
        \tikzstyle{node} = [draw, circle, thick]
        \node[type](titre) at (-1,0) {(d)};
        \node[node](n1) at (0,0) {$1$};
        \node[node](n2) at (1,-1.5) {$2$};
        \node[node](n3) at (-1,-1.5) {$3$};
        \draw[thick, ->] (n1) edge [bend left=20] node {$\geq$} (n2);
        \draw[LightSalmon4, thick, ->] (n2) edge [bend left=20] node {$<$} (n1);
        \draw[thick, ->] (n2) edge node {$\geq$} (n3);
		  \end{tikzpicture}}}
    \end{minipage}
    \hfill
    \begin{minipage}{.3\textwidth}
      \centerline{\scalebox{1}{\begin{tikzpicture}[>=latex,auto]
        \tikzstyle{type} = []
        \tikzstyle{node} = [draw, circle, thick]
        \node[type](titre) at (-1,0) {(e)};
        \node[node](n1) at (0,0) {$1$};
        \node[node](n2) at (1,-1.5) {$2$};
        \node[node](n3) at (-1,-1.5) {$3$};
        \draw[LightSalmon4, thick, ->] (n1) edge [bend left=20] node {$<$} (n2);
        \draw[thick, ->] (n2) edge [bend left=20] node {$\geq$} (n1);
        \draw[LightSalmon4, thick, ->] (n2) edge node {$<$} (n3);
		  \end{tikzpicture}}}      
    \end{minipage}
    \hfill
    \begin{minipage}{.3\textwidth}
      \centerline{\scalebox{1}{\begin{tikzpicture}[>=latex,auto]
        \tikzstyle{type} = []
        \tikzstyle{node} = [draw, circle, thick]
        \node[type](titre) at (-1,0) {(f)};
        \node[node](n1) at (0,0) {$1$};
        \node[node](n2) at (1,-1.5) {$2$};
        \node[node](n3) at (-1,-1.5) {$3$};
        \draw[thick, ->] (n1) edge [bend left=20] node {$\geq$} (n2);
        \draw[LightSalmon4, thick, ->] (n2) edge [bend left=20] node {$<$} (n1);
        \draw[LightSalmon4, thick, ->] (n2) edge node {$<$} (n3);
		  \end{tikzpicture}}}
    \end{minipage}\medskip
  \end{center}
  \caption{All the update digraphs associated to the digraph $G$ of Figure~\ref{fig:3digraph} at 
    stake in Example~\ref{ex:UD}, where 
    (a) is obtained with $s_1$ and $s_7$, 
    (b) is obtained with $s_2$, 
    (c) is obtained with $s_3$, $s_8$, $s_9$ and $s_{10}$, 
    (d) is obtained with $s_4$ and $s_{11}$, 
    (e) is obtained with $s_5$, and 
    (f) is obtained with $s_6$, $s_{12}$ and $s_{13}$.}
  \label{fig:UD}
\end{figure}
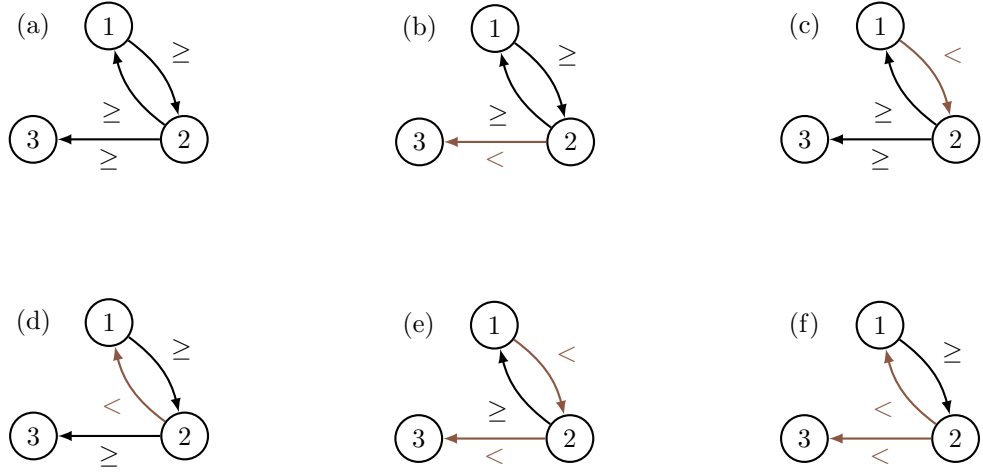

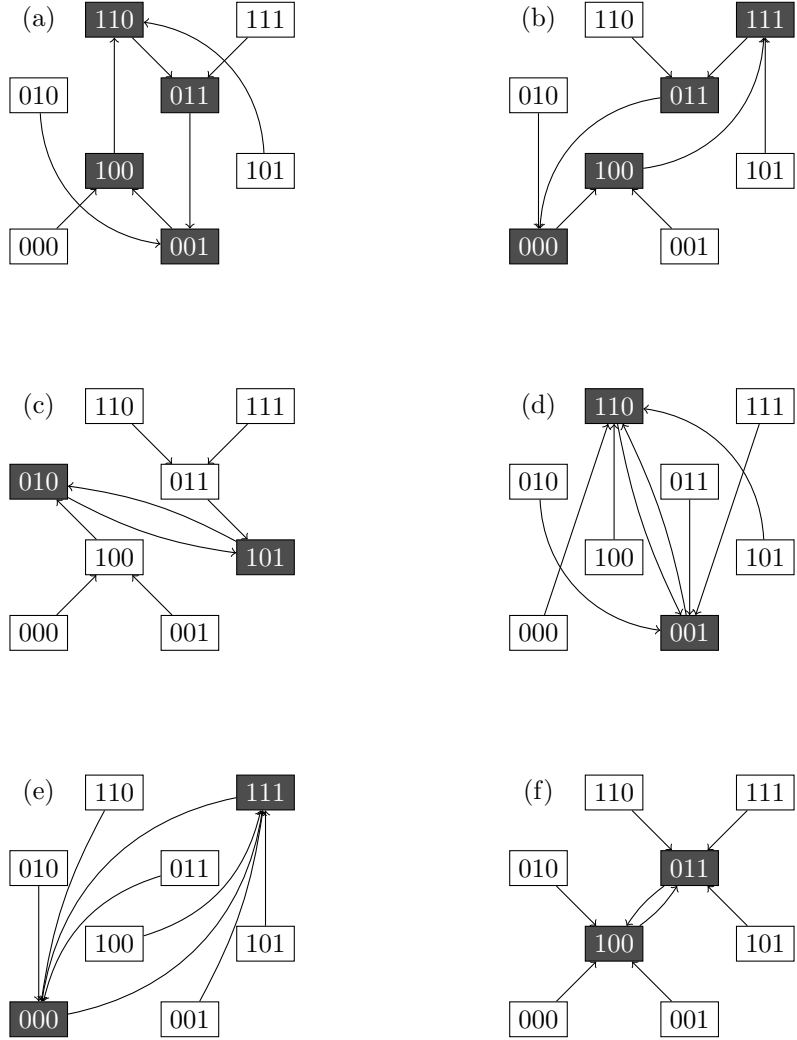
\begin{figure}[t!]
  	\begin{center}
    \begin{minipage}{.3\textwidth}
      \centerline{\scalebox{1}{\begin{tikzpicture}[>=to,auto]
        \tikzstyle{type} = []
        \tikzstyle{conf} = [rectangle, draw]
        \tikzstyle{pf} = [rectangle, draw, fill=black!10]
        \tikzstyle{lc} = [rectangle, draw, fill=black!70]
        \node[type](titre) at (0,3) {(a)};
        \node[conf](n000) at (0,0) {$000$};
        \node[lc](n001) at (2,0) {\textcolor{white}{$001$}};
        \node[conf](n010) at (0,2) {$010$};
        \node[lc](n011) at (2,2) {\textcolor{white}{$011$}};
        \node[lc](n100) at (1,1) {\textcolor{white}{$100$}};
        \node[conf](n101) at (3,1) {$101$};
        \node[lc](n110) at (1,3) {\textcolor{white}{$110$}};
        \node[conf](n111) at (3,3) {$111$};
        \draw[->] (n000) edge (n100);
        \draw[->] (n001) edge (n100);
        \draw[->] (n010) edge [bend right=40] (n001);
        \draw[->] (n011) edge (n001);
        \draw[->] (n100) edge (n110);
        \draw[->] (n101) edge [bend right=40] (n110);
        \draw[->] (n110) edge (n011);
        \draw[->] (n111) edge (n011);
      \end{tikzpicture}}}
    \end{minipage}
    \hspace*{20mm}
    \begin{minipage}{.3\textwidth}
      \centerline{\scalebox{1}{\begin{tikzpicture}[>=to,auto]
        \tikzstyle{type} = []
        \tikzstyle{conf} = [rectangle, draw]
        \tikzstyle{pf} = [rectangle, draw, fill=black!10]
        \tikzstyle{lc} = [rectangle, draw, fill=black!70]
        \node[type](titre) at (0,3) {(b)};
        \node[lc](n000) at (0,0) {\textcolor{white}{$000$}};
        \node[conf](n001) at (2,0) {$001$};
        \node[conf](n010) at (0,2) {$010$};
        \node[lc](n011) at (2,2) {\textcolor{white}{$011$}};
        \node[lc](n100) at (1,1) {\textcolor{white}{$100$}};
        \node[conf](n101) at (3,1) {$101$};
        \node[conf](n110) at (1,3) {$110$};
        \node[lc](n111) at (3,3) {\textcolor{white}{$111$}};
        \draw[->] (n000) edge (n100);
        \draw[->] (n001) edge (n100);
        \draw[->] (n010) edge (n000);
        \draw[->] (n011) edge [bend right=40] (n000);
        \draw[->] (n100) edge [bend right=40] (n111);
        \draw[->] (n101) edge (n111);
        \draw[->] (n110) edge (n011);
        \draw[->] (n111) edge (n011);
      \end{tikzpicture}}}
    \end{minipage}
  \end{center}\bigskip\bigskip
  
  \begin{center}
    \begin{minipage}{.3\textwidth}
      \centerline{\scalebox{1}{\begin{tikzpicture}[>=to,auto]
        \tikzstyle{type} = []
        \tikzstyle{conf} = [rectangle, draw]
        \tikzstyle{pf} = [rectangle, draw, fill=black!10]
        \tikzstyle{lc} = [rectangle, draw, fill=black!70]
        \node[type](titre) at (0,3) {(c)};
        \node[conf](n000) at (0,0) {$000$};
        \node[conf](n001) at (2,0) {$001$};
        \node[lc](n010) at (0,2) {\textcolor{white}{$010$}};
        \node[conf](n011) at (2,2) {$011$};
        \node[conf](n100) at (1,1) {$100$};
        \node[lc](n101) at (3,1) {\textcolor{white}{$101$}};
        \node[conf](n110) at (1,3) {$110$};
        \node[conf](n111) at (3,3) {$111$};
        \draw[->] (n000) edge (n100);
        \draw[->] (n001) edge (n100);
        \draw[->] (n010) edge [bend right=10] (n101);
        \draw[->] (n011) edge (n101);
        \draw[->] (n100) edge (n010);
        \draw[->] (n101) edge [bend right=10] (n010);
        \draw[->] (n110) edge (n011);
        \draw[->] (n111) edge (n011);
      \end{tikzpicture}}}
    \end{minipage}
    \hspace*{20mm}
    \begin{minipage}{.3\textwidth}
      \centerline{\scalebox{1}{\begin{tikzpicture}[>=to,auto]
        \tikzstyle{type} = []
        \tikzstyle{conf} = [rectangle, draw]
        \tikzstyle{pf} = [rectangle, draw, fill=black!10]
        \tikzstyle{lc} = [rectangle, draw, fill=black!70]
        \node[type](titre) at (0,3) {(d)};
        \node[conf](n000) at (0,0) {$000$};
        \node[lc](n001) at (2,0) {\textcolor{white}{$001$}};
        \node[conf](n010) at (0,2) {$010$};
        \node[conf](n011) at (2,2) {$011$};
        \node[conf](n100) at (1,1) {$100$};
        \node[conf](n101) at (3,1) {$101$};
        \node[lc](n110) at (1,3) {\textcolor{white}{$110$}};
        \node[conf](n111) at (3,3) {$111$};
        \draw[->] (n000) edge (n110);
        \draw[->] (n001) edge [bend right=7.5] (n110);
        \draw[->] (n010) edge [bend right=40] (n001);
        \draw[->] (n011) edge (n001);
        \draw[->] (n100) edge (n110);
        \draw[->] (n101) edge [bend right=40] (n110);
        \draw[->] (n110) edge [bend right=7.5] (n001);
        \draw[->] (n111) edge (n001);
      \end{tikzpicture}}}
    \end{minipage}
  \end{center}\bigskip\bigskip
    
  \begin{center}
    \begin{minipage}{.3\textwidth}
      \centerline{\scalebox{1}{\begin{tikzpicture}[>=to,auto]
        \tikzstyle{type} = []
        \tikzstyle{conf} = [rectangle, draw]
        \tikzstyle{pf} = [rectangle, draw, fill=black!10]
        \tikzstyle{lc} = [rectangle, draw, fill=black!70]
        \node[type](titre) at (0,3) {(e)};
        \node[lc](n000) at (0,0) {\textcolor{white}{$000$}};
        \node[conf](n001) at (2,0) {$001$};
        \node[conf](n010) at (0,2) {$010$};
        \node[conf](n011) at (2,2) {$011$};
        \node[conf](n100) at (1,1) {$100$};
        \node[conf](n101) at (3,1) {$101$};
        \node[conf](n110) at (1,3) {$110$};
        \node[lc](n111) at (3,3) {\textcolor{white}{$111$}};
        \draw[->] (n000) edge [bend right=35] (n111);
        \draw[->] (n001) edge [bend right=10] (n111);
        \draw[->] (n010) edge  (n000);
        \draw[->] (n011) edge [bend right=30] (n000);
        \draw[->] (n100) edge [bend right=30] (n111);
        \draw[->] (n101) edge (n111);
        \draw[->] (n110) edge [bend right=10] (n000);
        \draw[->] (n111) edge [bend right=35] (n000);
      \end{tikzpicture}}}
    \end{minipage}
    \hspace*{20mm}
    \begin{minipage}{.3\textwidth}
      \centerline{\scalebox{1}{\begin{tikzpicture}[>=to,auto]
        \tikzstyle{type} = []
        \tikzstyle{conf} = [rectangle, draw]
        \tikzstyle{pf} = [rectangle, draw, fill=black!10]
        \tikzstyle{lc} = [rectangle, draw, fill=black!70]
        \node[type](titre) at (0,3) {(f)};
        \node[conf](n000) at (0,0) {$000$};
        \node[conf](n001) at (2,0) {$001$};
        \node[conf](n010) at (0,2) {$010$};
        \node[lc](n011) at (2,2) {\textcolor{white}{$011$}};
        \node[lc](n100) at (1,1) {\textcolor{white}{$100$}};
        \node[conf](n101) at (3,1) {$101$};
        \node[conf](n110) at (1,3) {$110$};
        \node[conf](n111) at (3,3) {$111$};
        \draw[->] (n000) edge (n100);
        \draw[->] (n001) edge (n100);
        \draw[->] (n010) edge (n100);
        \draw[->] (n011) edge [bend right=10] (n100);
        \draw[->] (n100) edge [bend right=10] (n011);
        \draw[->] (n101) edge (n011);
        \draw[->] (n110) edge (n011);
        \draw[->] (n111) edge (n011);
      \end{tikzpicture}}}
    \end{minipage}
  \end{center}
  \caption{The six distinct possible block-sequential dynamics of the Boolean automata network of 
    Example~\ref{ex:UD}, where (a), (b), (c), (d), (e) and (f) respectively derive from update modes 
    equivalence classes $[s_1]_G$, $[s_2]_G$, $[s_4]_G$, $[s_5]_G$, $[s_6]_G$ and $[s_8]_G$ defined 
    in Table~\ref{tab:UDeq}.}
  \label{fig:UDdynamics}
\end{figure}

\begin{example}\normalsize\normalfont\label{ex:UD}
  This example aims to illustrate all the previous concepts and results in detail.
  Consider BAN$_3$ $f$ defined in Figure~\ref{fig:3digraph}(a) and its interaction graph depicted in 
  Figure~\ref{fig:3digraph}(b). 
  What follows can be easily verified:
  \begin{enumerate}
  \item There exist exactly $|\text{BS}_n| = 13$ block-sequential update modes: 
    $s_1 \equiv (\{1, 2, 3\})$, 
    $s_2 \equiv (\{1, 2\}, \{3\})$, 
    $s_3 \equiv (\{1, 3\}, \{2\})$, 
    $s_4 \equiv (\{2, 3\},\{1\})$, 
    $s_5 \equiv (\{1\},\{2, 3\})$, 
    $s_6 \equiv (\{2\},\{1, 3\})$, 
    $s_7 \equiv (\{3\},\{1, 2\})$, 
    $s_8 \equiv (\{1\},\{2\},\{3\})$, 
    $s_9 \equiv (\{1\},\{3\},\{2\})$, 
    $s_{10} \equiv (\{2\},\{1\},\{3\})$, 
    $s_{11} \equiv (\{2\},\{3\},\{1\})$, 
    $s_{12} \equiv (\{3\},\{1\},\{2\})$, 
    $s_{13} \equiv (\{3\},\{2\},\{1\})$, 
    $s_1$ being the parallel mode.
  \item There are $8$ possible labeled digraphs (two possible labels for each edge). 
    From them, only the $6$ labeled digraphs shown in Figure~\ref{fig:UD} satisfy 
    Theorem~\ref{thm:arac2011}. 
    This means that some of the update digraphs generated by the $13$ previous update modes are 
    necessarily equal, which leads to produce the $6$ equivalence classes presented in 
    Table~\ref{tab:UDeq}.
  \end{enumerate}
  Therefore, if one wishes to exhaustively study all the distinct dynamics that a BAN$_3$ with its 
  associated digraph can have, it is not necessary to use all $13$ possible update modes. 
  Instead, only six of them are needed (a reduction of 54\%), which are representative in the sense 
  that they define all equivalence classes of update modes with respect to update digraphs. 
  Consequently, such a BAN$_3$ will have at most six distinct dynamics by 
  Theorem~\ref{thm:arac2009}. 
  Figures~\ref{fig:3digraph} to~\ref{fig:UDdynamics} show a concrete example of all the different 
  block-sequential dynamics that BAN$_3$ $f$ with interaction graph $G_f$ can have. 
\end{example}

{\small
\begin{table}[b!]
  \centerline{
    \begin{tabular}{@{}llllll@{}}
      \toprule
      $[s_1]_G$ & $[s_2]_G$ & $[s_4]_G$ & $[s_5]_G$ & $[s_6]_G$ & $[s_8]_G$\\
      \midrule
      $(\{1, 2, 3\})$ & $(\{1, 2\}, \{3\})$ & $(\{2, 3\},\{1\})$ & $(\{1\}, \{2, 3\})$ & 
      $(\{2\}, \{1, 3\})$ & $(\{1\},\{2\},\{3\})$\\
      $(\{3\},\{1, 2\})$  &  & $(\{3\},\{2\},\{1\})$ & $(\{1, 3\}, \{2\})$ &  & 
      $(\{2\},\{1\},\{3\})$\\
      &  &  & $(\{1\},\{3\},\{2\})$ &  & $(\{2\},\{3\},\{1\})$\\
      &  &  & $(\{3\},\{1\},\{2\})$ &  & \\
      \bottomrule
    \end{tabular}
  }
  \caption{The different equivalence classes associated to $G$ from Figure~\ref{fig:3digraph} and 
    the updating modes contained in each one. Representative modes of each class are the ones which 
    appear in the first line of each column.}
  \label{tab:UDeq}
\end{table}
}

Note that having the information of all the different dynamics, one can discuss any dynamical 
property of the network, for instance, robustness, which could be understood in the sense of whether 
the observed dynamics retain certain attractors. 
In this sense, the BAN of Example~\ref{ex:UD} is not robust. 
In the theory of update digraphs developed, other properties are also proven which justify the fact 
that this reduction in the number of modes needed to study the distinct dynamics of a BAN is 
generally much greater than the 54\% reduction observed in this example, as the network becomes 
larger.
In this way, we will see what happens beyond the parallel update mode which could be 
biased~\cite{Harvey1997} or not robust against perturbations of the update modes as occurs in 
certain families of BANs such as elementary cellular automata, where it is possible to 
find very robust networks called block invariant and other ones that are 
not~\cite{Goles2015,Goles2018,Perrot2020,Ruivo2020}.\medskip

\begin{algorithm}[t!]
  \textbf{Input:} An update digraph $(G, \text{lab})$.\\
  \textbf{Output:} The most compact update mode associated to $(G, \text{lab})$.
  \medskip
  \hrule
  \begin{enumerate}
  \item[1)] From $(G, \text{lab})$, construct $(G, \text{lab})^{[\geq]}$ in which all the 
    vertices belonging to a same strongly connected component composed only of $\geq$-edges 
    are merged into a unique representative vertex. 
  \item[2)] From $(G, \text{lab})^{[\geq]}$, construct $G' = (G, \text{lab})_\text{rev}^{[\geq]}$ 
    in which all $<$-edge is reversed thanks to a parallel procedure on every edge of 
    $(G, \text{lab})^{[\geq]}$ transforming all $<$-edge $(i,j)$ into the $<$-edge 
    $(j, i)$.
  \item[3)] $t := 0$. \hfill \textsf{\# $t=0$ is a substep corresponding to the construction of 
    block $W_1$.}
  \item[4)] Compute on $G'$ the set of paths $P_<$ composed of all the longest paths with the 
    maximum number of $<$-edges such that $P_< = \{P \mid \#(<\text{-edge} \in P)
    \text{ is max.}\}$.\\ 
    If $P_< = \emptyset$, then goto 7).
  \item[5)] Let $T$ be the set of all target vertices of the last $<$-edge of each $P$ of 
    $P_<$, and let $S(T)$ the set of all the successors of elements of $T$ in $G'$. Every elements 
    of $T$ and $S(T)$ are scheduled simultaneously at substep $t$.\\
    $t := t + 1$.
  \item[6)] Remove all elements of $T$, $S(T)$ and all their incoming edges from $G'$, and go back 
    to 4).
  \item[7)] All the remaining vertices are scheduled simultaneously at substep $t$.
  \end{enumerate}
  \caption{Compact representative block-sequential update modes.}
  \label{algo:compactUM}
\end{algorithm}

Note that in~\cite{Aracena2013a}, an algorithm was provided to determine representative modes for 
each equivalent class. 
We present in Algorithm~\ref{algo:compactUM} a new polynomial algorithm which computes the most 
compact representative block-sequential update mode in terms of the dimension of its ordered 
partition, given an update digraph $(G, \text{lab}_s)$.

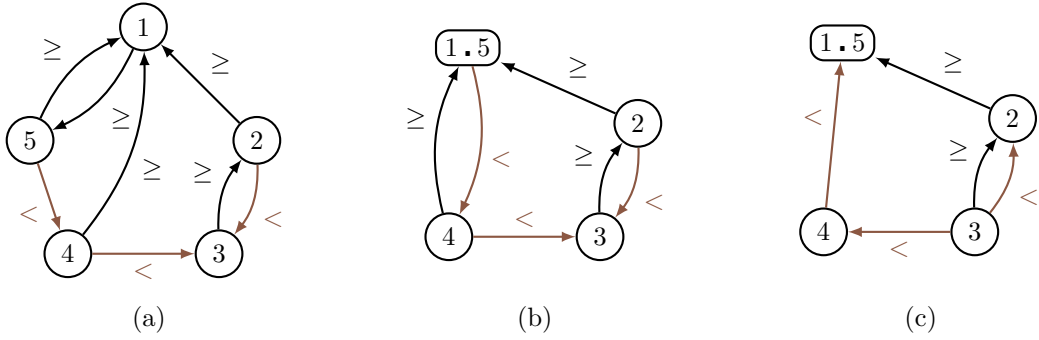
\begin{figure}[t!]
  \begin{center}
    \begin{minipage}{.3\textwidth}
			\centerline{\scalebox{1}{\begin{tikzpicture}[>=latex,auto]
				\tikzstyle{node} = [draw, circle, thick]
				\tikzstyle{merged} = [draw, rounded corners=5pt, thick]
				\node[node](n1) at (0,0) {$1$};
				\node[node](n2) at (1.5,-1.5) {$2$};
				\node[node](n3) at (1,-3) {$3$};
				\node[node](n4) at (-1,-3) {$4$};
				\node[node](n5) at (-1.5,-1.5) {$5$};
				\draw[thick, ->] (n1) edge [bend left=20] node {$\geq$} (n5);
				\draw[thick, ->] (n2) edge node [swap] {$\geq$} (n1);
				\draw[LightSalmon4, thick, ->] (n2) edge [bend left=20] node {$<$} (n3);
				\draw[thick, ->] (n3) edge [bend left=20] node {$\geq$} (n2);
				\draw[thick, ->] (n4) edge [bend right=20] node [swap] {$\geq$} (n1);
				\draw[LightSalmon4, thick, ->] (n4) edge node [swap] {$<$} (n3);
				\draw[thick, ->] (n5) edge [bend left=20] node {$\geq$} (n1);
				\draw[LightSalmon4, thick, ->] (n5) edge node [swap] {$<$} (n4);
			\end{tikzpicture}}}
		\end{minipage}
		\hfill
		\begin{minipage}{.3\textwidth}
			\centerline{\scalebox{1}{\begin{tikzpicture}[>=latex,auto]
				\tikzstyle{node} = [draw, circle, thick]
				\tikzstyle{merged} = [draw, rounded corners=5pt, thick]
				\node[merged](n15) at (-0.75,-0.5) {$1\centerdot{}5$};
				\node[node](n2) at (1.5,-1.5) {$2$};
				\node[node](n3) at (1,-3) {$3$};
				\node[node](n4) at (-1,-3) {$4$};
				\draw[LightSalmon4, thick, ->] (n15) edge [bend left=20] node {$<$} (n4);
				\draw[thick, ->] (n2) edge node [swap] {$\geq$} (n15);
				\draw[LightSalmon4, thick, ->] (n2) edge [bend left=20] node {$<$} (n3);
				\draw[thick, ->] (n3) edge [bend left=20] node {$\geq$} (n2);
				\draw[thick, ->] (n4) edge [bend left=20] node {$\geq$} (n15);
				\draw[LightSalmon4, thick, ->] (n4) edge node {$<$} (n3);
			\end{tikzpicture}}}
		\end{minipage}
		\hfill
		\begin{minipage}{.3\textwidth}
			\centerline{\scalebox{1}{\begin{tikzpicture}[>=latex,auto]
				\tikzstyle{node} = [draw, circle, thick]
				\tikzstyle{merged} = [draw, rounded corners=5pt, thick]
				\node[merged](n15) at (-0.75,-0.5) {$1\centerdot{}5$};
				\node[node](n2) at (1.5,-1.5) {$2$};
				\node[node](n3) at (1,-3) {$3$};
				\node[node](n4) at (-1,-3) {$4$};
				\draw[thick, ->] (n2) edge node [swap] {$\geq$} (n15);
				\draw[LightSalmon4, thick, ->] (n3) edge [bend right=20] node [swap] {$<$} (n2);
				\draw[thick, ->] (n3) edge [bend left=20] node {$\geq$} (n2);
				\draw[LightSalmon4, thick, ->] (n3) edge node {$<$} (n4);
				\draw[LightSalmon4, thick, ->] (n4) edge node {$<$} (n15);
			\end{tikzpicture}}}
		\end{minipage}\medskip
      
		\begin{minipage}{.3\textwidth}
			\centerline{(a)}
		\end{minipage}
		\hfill
		\begin{minipage}{.3\textwidth}
			\centerline{(b)}
		\end{minipage}
		\hfill
		\begin{minipage}{.3\textwidth}
			\centerline{(c)}
		\end{minipage}
  \end{center}
  \caption{(a) An update digraph $(G, \text{lab})$; 
    (b) its related quotient $(G, \text{lab})^{[\geq]}$ according to the equivalence relation ``is 
    accessible through $\geq$-edges'' (cf. Algorithm); 
    (c) $(G, \text{lab})_{\text{rev}}^{[\geq]}$ obtained from (b) by reversing all the $<$-edges.}
  \label{fig:compactUM}
\end{figure}

\begin{example}\normalsize\normalfont\label{ex:algo-compactUM}
  This example aims to clarify the construction of a block-sequential update mode among the 
  most compact as it is defined in Algorithm~\ref{algo:compactUM}. 
  To do so, consider the labeled digraph depicted $(G, \text{lab})$ in 
  Figure~\ref{fig:compactUM}(a). 
  It is composed of five vertices and eight edges, among which five are $\geq$-labeled and three are 
  $<$-labeled. 
  Step 1) of the algorithm makes us compute that there exists a unique non-trivial strongly 
  connected component composed only of $\geq$-edges. 
  This component is the subgraph composed of vertices 1 and 5.
  These two vertices can be merged into a representative since they act the same way in the building 
  process of the expected block-sequential update mode, which induces that it is useless to conserve 
  both of them. 
  So, step 1) leads to merge these two vertices and build $(G, \text{lab})^{[\geq]}$ depicted in 
  Figure~\ref{fig:compactUM}(b). 
  From $(G, \text{lab})^{[\geq]}$, executing step 2) aims at reversing all the $<$-edges, 
  which leads to $G' = (G, \text{lab})_\text{rev}^{[\geq]}$ depicted in 
  Figure~\ref{fig:compactUM}(c).
  This is the labeled digraph on which the core of algorithm acts. 
  First, notice that this graph does not admit any cycle composed of at least one $<$-edge. 
  Although the input specifies that $(G, \text{lab})$ is an update digraph, it is interesting to 
  notice that Theorem~\ref{thm:arac2011} guaranties at this stage that $G'$ is indeed an update 
  digraph. 
  Step 3) affects value $0$ to variable $t$, which will help us to build the first block ($W_1$) of 
  the desired update mode. 
  At step 4), we compute the set $P_<$ of all the longest paths in $G'$ which contains the maximum 
  of $<$-edges. In $G'$, there is a unique such path composed of two $<$-edges and 
  $P_< = \{(3, 4, 1\centerdot{}5)\}$. Since $P \neq \emptyset$, the following step is step 5). 
  Step 5) asks to compute the target vertices of the last $<$-edge of each element of 
  $P_<$. 
  In our case, there is one such target vertex, $1\centerdot{}5$. 
  This vertex does not have any successor. 
  So, $T = \{1\centerdot{}5\}$ and $S(T) = \emptyset$. 
  Thus, now we know that the first block of the expected update mode is $W_1 = \{1,5\}$ since 
  $1\centerdot{}5$ is the represented of both initial vertices $1$ and $5$. 
  The value of $t$ is then incremented and equals $1$ at the end of step 5). 
  Executing step 6) leads to remove vertex $1\centerdot{}5$ and all its incoming edges from $G'$ 
  such that $G'$ becomes the graph composed of vertices $2$, $3$ and $4$ and of edges $(3, <, 2)$ 
  and $(3, <, 4)$. This steps ends with a goto instruction to step 4). 
  Executing step 4) at this stage computes $P_< = \{(3, 2), (3, 4)\}$, with two paths composed of a 
  unique $<$-edges.
  Executing step 5) computes $T = \{2, 4\}$ and $S(T) = \emptyset$ and makes adding $W_2 = \{2,4\}$ 
  as the second block in the update mode in construction. 
  The value of $t$ is then incremented and equals $2$. 
  Step 6) leads to remove vertices $2$ and $4$ and their incoming edges from $G'$ such the latter 
  is now composed of only vertex $3$ and no edge.
  Then, we go back to step 4). 
  At this stage, since $G'$ has no edge, there is no path and $P_< = \emptyset$, which leads to step 
  7), which makes adding $W_3 = \{3\}$ as the third and last block in the update mode in 
  construction. 
  The execution of the algorithm returns $\mu = (\{1,5\}, \{2,4\}, \{3\})$.\medskip
\end{example}

It is easy to show that Algorithm~\ref{algo:compactUM} is correct. 
Indeed, its output is necessarily a block-sequential update mode among the most compact admissible 
ones since, by construction, it builds as much blocks as the maximum number of reversed $<$-labeled 
plus $1$ in a path of $G'$, which is optimal since two vertices $i$ and $j$ which are connected to 
each other through a $<$-edge necessarily belong to two distinct blocks by definition.\medskip

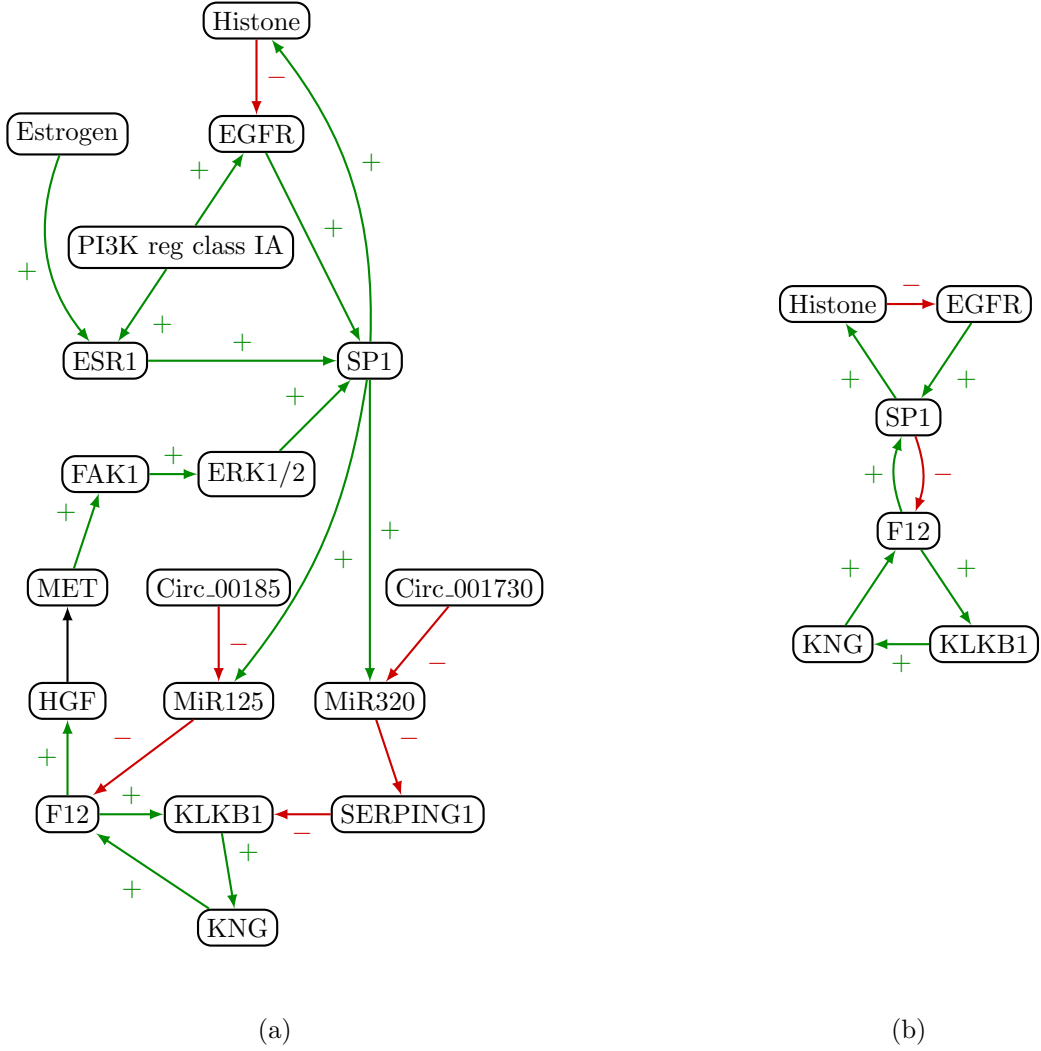
\begin{figure}[t!]
  \begin{center}
    \begin{minipage}{.55\textwidth}
			\centerline{\scalebox{1}{\begin{tikzpicture}[>=latex,auto]
        \tikzstyle{node} = [draw, rounded corners=5pt, thick]
        \node[node](nEGFR) at (0,0) {EGFR};
        \node[node](nPI3K1) at (-1,-1.5) {PI3K reg class IA};
        \node[node](nESR1) at (-2,-3) {ESR1};
        \node[node](nEstrogen) at (-2.5,0) {Estrogen};
        \node[node](nHistone) at (0,1.5) {Histone};
        \node[node](nSP1) at (1.5,-3) {SP1};
        \node[node](nERK1/2) at (0,-4.5) {ERK1/2};
        \node[node](nFAK1) at (-2,-4.5) {FAK1};
        \node[node](nCirc00185) at (-.5,-6) {Circ\_00185};
        \node[node](nCirc001730) at (2.75,-6) {Circ\_001730};
        \node[node](nMiR125) at (-0.5,-7.5) {MiR125};
        \node[node](nMiR320) at (1.5,-7.5) {MiR320};        
        \node[node](nF12) at (-2.5,-9) {F12};
        \node[node](nHGF) at (-2.5,-7.5) {HGF};
        \node[node](nMET) at (-2.5,-6) {MET};
        \node[node](nSERPING1) at (2,-9) {SERPING1};
        \node[node](nKLKB1) at (-0.5,-9) {KLKB1};
        \node[node](nKNG) at (-.25,-10.5) {KNG};
        
				\draw[Green4, thick, ->] (nEGFR) edge node {$+$} (nSP1);
				\draw[Green4, thick, ->] (nPI3K1) edge node {$+$} (nEGFR);
				\draw[Green4, thick, ->] (nPI3K1) edge node {$+$} (nESR1);
				\draw[Green4, thick, ->] (nESR1) edge node {$+$} (nSP1);
				\draw[Green4, thick, ->] (nEstrogen) edge [bend right=30] node [swap] {$+$} (nESR1);
				\draw[Red3, thick, ->] (nHistone) edge node {$-$} (nEGFR);
				\draw[Green4, thick, ->] (nSP1) edge [bend right=20] node [swap]{$+$} (nHistone);
				\draw[Green4, thick, ->] (nSP1) edge [bend left=15] node {$+$} (nMiR125);
				\draw[Green4, thick, ->] (nSP1) edge node {$+$} (nMiR320);
				\draw[Green4, thick, ->] (nERK1/2) edge node {$+$} (nSP1);
				\draw[Green4, thick, ->] (nFAK1) edge node {$+$} (nERK1/2);
				\draw[Red3, thick, ->] (nCirc00185) edge node {$-$} (nMiR125);
				\draw[Red3, thick, ->] (nCirc001730) edge node {$-$} (nMiR320);
				\draw[Red3, thick, ->] (nMiR125) edge node [swap] {$-$} (nF12);
				\draw[Red3, thick, ->] (nMiR320) edge node {$-$} (nSERPING1);
				\draw[Green4, thick, ->] (nF12) edge node {$+$} (nHGF);
				\draw[Green4, thick, ->] (nF12) edge node {$+$} (nKLKB1);
				\draw[thick, ->] (nHGF) edge (nMET);
				\draw[Green4, thick, ->] (nMET) edge node {$+$} (nFAK1);
				\draw[Red3, thick, ->] (nSERPING1) edge node {$-$} (nKLKB1);
				\draw[Green4, thick, ->] (nKLKB1) edge node {$+$} (nKNG);
				\draw[Green4, thick, ->] (nKNG) edge node {$+$} (nF12);
			\end{tikzpicture}}}
		\end{minipage}
		\hfill
		\begin{minipage}{.3\textwidth}
			\centerline{\scalebox{1}{\begin{tikzpicture}[>=latex,auto]
        \tikzstyle{node} = [draw, rounded corners=5pt, thick]
        \node[node](nEGFR) at (0,0) {EGFR};
        \node[node](nHistone) at (-2,0) {Histone};
        \node[node](nSP1) at (-1,-1.5) {SP1};
        \node[node](nF12) at (-1,-3) {F12};
        \node[node](nKLKB1) at (0,-4.5) {KLKB1};
        \node[node](nKNG) at (-2,-4.5) {KNG};
        
				\draw[Red3, thick, ->] (nHistone) edge node {$-$} (nEGFR);
				\draw[Green4, thick, ->] (nEGFR) edge node {$+$} (nSP1);
				\draw[Green4, thick, ->] (nSP1) edge node {$+$} (nHistone);
				\draw[Red3, thick, ->] (nSP1) edge [bend left=20] node {$-$} (nF12);
				\draw[Green4, thick, ->] (nF12) edge [bend left=20] node {$+$} (nSP1);
				\draw[Green4, thick, ->] (nF12) edge node {$+$} (nKLKB1);
				\draw[Green4, thick, ->] (nKLKB1) edge node {$+$} (nKNG);
				\draw[Green4, thick, ->] (nKNG) edge node {$+$} (nF12);
			\end{tikzpicture}}}
		\end{minipage}
  \end{center}\medskip
  
  \begin{center}
    \begin{minipage}{.55\textwidth}
      \centerline{(a)}
    \end{minipage}
    \hfill
    \begin{minipage}{.3\textwidth}
      \centerline{(b)}
    \end{minipage}
  \end{center}
  \caption{(a) Genetic regulatory network integrating different genes and proteins involved in the 
    hereditary angioedema, in which the black arrow between HGF and MET corresponds to a signalling 
    pathway; (b) Simplified version of this network retaining only the major players 
    in the regulation of gene expression.}
  \label{fig:GRN}
\end{figure}

Moreover it is easy to show that Algorithm~\ref{algo:compactUM} time complexity is polynomial 
according to the number of vertices of the graph. 
Let us consider that $(G, \text{lab})$ is composed of $n$ vertices.
Then the following holds:
\begin{itemize}
\item[--] Step 1) needs to compute the strongly connected components of $G$, which can be done in 
  $O(n^2)$~\cite{Tarjan1972}. Then, we have to choose among these components only those which are 
  non-trivial and composed only of $\geq$-edges to construct the quotient graph. It can be 
  done in $O(n^2)$ too. 
\item[--] Step 2) can be done in $O(n^2)$ since it consists in reversing all the edges in the worst 
  case. 
\item[--] Step 3) is trivially in $O(1)$. 
\item[--] Step 4) can be done in $O(n^2)$. Indeed, $G'$ is necessarily acyclic since all the cycles 
  composed only of $\geq$-edges are removed by step 1) and, given that $(G, \text{lab})$ is 
  an update digraph, by Theorem~\ref{thm:arac2011}, $G'$ is necessarily cycle-free at the end of   
  step 2). 
  Thus, step 4) consists in computing a topological ordering of $G'$ for which classical 
  algorithms are in $O(n^2)$~\cite{Cormen2001}, and then, searching for the longest paths 
  maximizing the number of $<$-edges thanks to a depth-first search approach from the nodes of the 
  first layer of the ordering. 
  This can also be done in $O(n^2)$.
\item[--] The part of step 5) which consists in computing the sets $T$ and $S(T)$ can be found 
  directly from step 4) and is not significant in terms of time complexity. 
  Then, constructing the block can be done in $O(n)$ and incrementing the value of $t$ is done in 
  constant time.
\item[--] Step 6) consists only in removing vertices and their incoming arcs from $G'$. It can be 
  done in $O(n)$.
\item[--] Step 7) constructs the last block, which can be done in $O(n)$. 
\end{itemize}
Eventually, steps 4) to 6) are repeated a number of times equal to the maximum number of $<$-edges 
in a path, which is majored by the total number of $<$-edges of $G'$. 
Hence, the algorithm is polynomial in $n$.

\section{Results and discussion}\label{sec:results}

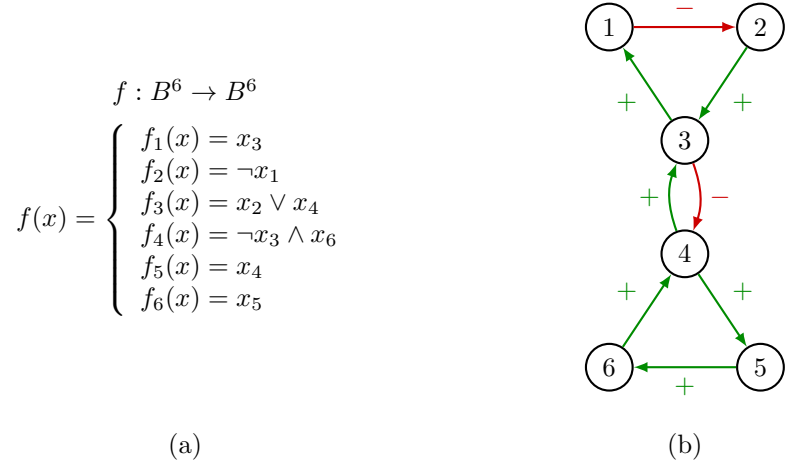
\begin{figure}[t!]
  \begin{center}
    \begin{minipage}{.3\textwidth}
      \centerline{$f: B^6 \to B^6$}\medskip
			
      \centerline{
        $f(x) = \left\{\begin{array}{l}
          f_1(x) = x_3\\
          f_2(x) = \neg x_1\\
          f_3(x) = x_2 \lor x_4\\
          f_4(x) = \neg x_3 \land x_6\\
          f_5(x) = x_4\\
          f_6(x) = x_5\\
        \end{array}\right.$
      }      
    \end{minipage}
    \hspace*{20mm}
    \begin{minipage}{.3\textwidth}
      	\centerline{\scalebox{1}{\begin{tikzpicture}[>=latex,auto]
        \tikzstyle{node} = [draw, circle, thick]
        \node[node](n1) at (-2,0) {1};
        \node[node](n2) at (0,0) {2};
        \node[node](n3) at (-1,-1.5) {3};
        \node[node](n4) at (-1,-3) {4};
        \node[node](n5) at (0,-4.5) {5};
        \node[node](n6) at (-2,-4.5) {6};
        
				\draw[Red3, thick, ->] (n1) edge node {$-$} (n2);
				\draw[Green4, thick, ->] (n2) edge node {$+$} (n3);
				\draw[Green4, thick, ->] (n3) edge node {$+$} (n1);
				\draw[Red3, thick, ->] (n3) edge [bend left=20] node {$-$} (n4);
				\draw[Green4, thick, ->] (n4) edge [bend left=20] node {$+$} (n3);
				\draw[Green4, thick, ->] (n4) edge node {$+$} (n5);
				\draw[Green4, thick, ->] (n5) edge node {$+$} (n6);
				\draw[Green4, thick, ->] (n6) edge node {$+$} (n4);
			\end{tikzpicture}}}
    \end{minipage}
  \end{center}

  \begin{center}
    \begin{minipage}{.3\textwidth}
      \centerline{(a)}
    \end{minipage}
    \hspace*{20mm}
    \begin{minipage}{.3\textwidth}
      \centerline{(b)}
    \end{minipage}
  \end{center}
  \caption{(a) The Boolean automata network associated with the simplified version of the genetic 
    regulatory network of hereditary angioedema; (b) Its associated interaction graph.}
  \label{fig:BAN-GRN}
\end{figure}

\subsection{Genetic regulatory network of hereditary angioedema}\label{sec:results:example}

Figure~\ref{fig:GRN} pre\-sents a genetic regulatory network regulating the genes involved in 
hereditary angioedema (note: This network is derived from an extended version from the literature of 
the one initially Clarivate MetaCore$^{\text{\textregistered}}$), a familial disease with a 
prevalence of about 1/50{,}000, with two types of expression of SERPING1, one with low (type I) and 
another with high (type II) expression level~\cite{Rachdi2020,Vincent2024}. 
Boolean automata networks which model such a disease needs to admit dynamics with at least $3$ 
attractors, to represent the physiologic and the two pathologic behaviors corresponding to different 
profiles of gene expression.

\subsection{Robustness to block-sequential update modes}\label{sec:results:robustness}

\begin{figure}[t!]
  \begin{center}
    \begin{minipage}[t!]{.22\textwidth}
      \centerline{\includegraphics[scale=.33]{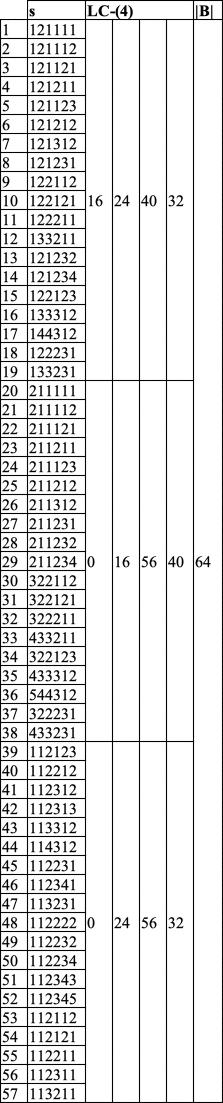}}
    \end{minipage}
    \hfill
    \begin{minipage}[t!]{.165\textwidth}
      \centerline{\includegraphics[scale=.33]{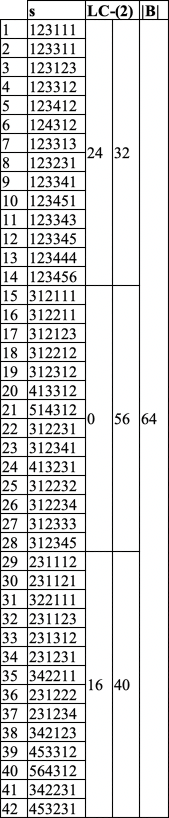}}
    \end{minipage}
    \hfill
    \begin{minipage}[t!]{.265\textwidth}
      \centerline{\includegraphics[scale=.33]{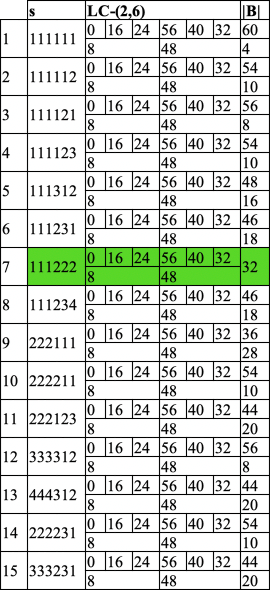}}\bigskip

      \centerline{\includegraphics[scale=.33]{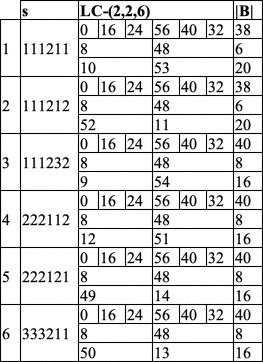}}
    \end{minipage}
    \hfill
    \begin{minipage}[t!]{.21\textwidth}
      \centerline{\includegraphics[scale=.33]{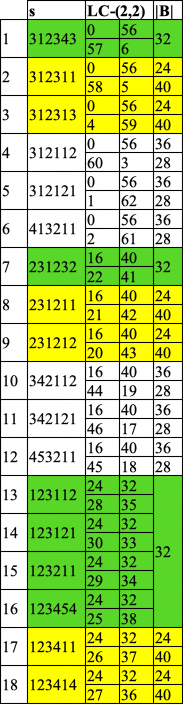}}\bigskip
      
      \centerline{\includegraphics[scale=.33]{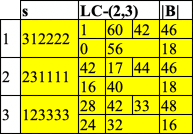}}\bigskip
      
      \centerline{\includegraphics[scale=.33]{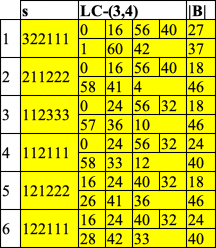}}
    \end{minipage}
  \end{center}
  \caption{All the possible block-sequential attractors of Boolean automata network $\Psi$ defined 
    in Figure~\ref{fig:BAN-GRN}. 
  }
  \label{fig:BAN-UDdyn}
\end{figure}

Let us consider from now on the BAN$_6$ $\Psi$ defined in Figure~\ref{fig:BAN-GRN} which is the 
BAN whose dynamics is the same as that of the following TBAN defined by its interaction matrix $w$ 
and its activation threshold vector $\theta$:
\begin{equation*}
  w = \left(\begin{array}{cccccc}
    0  & 0 & 1  & 0 & 0 & 0\\
    -1 & 0 & 0  & 0 & 0 & 0\\
    0  & 1 & 0  & 1 & 0 & 0\\
    0  & 0 & -1 & 0 & 0 & 1\\
    0  & 0 & 0  & 1 & 0 & 0\\
    0  & 0 & 0  & 0 & 1 & 0\\
  \end{array}\right) \quad \text{and} \quad 
  \theta = \begin{pmatrix}
    0\\
    0\\
    0\\
    -\epsilon < 0\\
    0\\
    0\\
  \end{pmatrix}
\end{equation*}
which is directly based on the simplified version of the genetic regulatory network described 
in~\ref{fig:GRN}, and whose interaction weights and activation thresholds have been chosen in 
coherence with the data of the literature. 
This network has the following properties:
\begin{enumerate}
\item The total number of underlying block-sequential update modes is $|BS_6| = 4683$.
\item It is easy to check that in a network $f$ whose interaction graph is a chain of cycles   
  (\emph{i.e.} a topological sequence of cycles such that two neighbor cycles intersect in a unique 
  vertex) composed of $k$ connected cycles of length $p_i$, $i \in \{1, \dots, k\}$, the number 
  of its update digraphs is given by $|U(f)| = \prod_{i=1}^{k} (2p_i - 1)$. 
  In the case of $\Psi$, since its interaction graph is a particular chain of $3$ cycles, with 
  $k = 3$, $p_1 = p_3 = 3$ and $p_2 = 2$, then $|U(\Psi)| = (2^3 - 1)^2 \cdot (2^2 - 1) = 147$. 
  This means that network $\Psi$ has at most $147$ different block-se\-quen\-tial dynamics within 
  the 
  $4683$ possible ones ($3.1\%$ of $4683$). 
  As mentioned in Section~\ref{sec:context}, one way to obtain these representative update modes 
  consists in using either the algorithm proposed in~\cite{Aracena2013a} or 
  Algorithm~\ref{algo:compactUM}.
\item The results of exhaustive simulations over the $147$ representative block-se\-quen\-tial 
  update modes is presented in Figure~\ref{fig:BAN-UDdyn}. 
  These results emphasize the limit cycles which are reached by these $147$ block-sequential 
  dynamics, and the sizes of their related attraction basins. 
  In this figure, we have changed the update mode notation to make it more compact. 
  Given a block-sequential update mode $\mu = (W_i)_{i \in M = \{1, \dots, m\}, m \leq n}$ and its 
  rewriting as function $s : N \to M$, it is represented by a vector of dimension $n$ (encoded as a 
  word over the alphabet $M$) in which each position represents an automaton whose associated 
  value gives the time substep (of the period induced by the update mode) at which the node is 
  updated.
  Furthermore, Boolean configurations are represented by their base-ten encoding for the sake of 
  concision. 
  For instance, the row $36$ of Table LC-$(4)$ means that for update mode $s = 544312$ 
  (corresponding to $\mu = (\{5\},\{6\},\{4\},\{2,3\},\{1\})$), the associated dynamics generated 
  only one limit cycle of length $4$ (and therefore with an attraction basin of maximum size 
  $|B| = 2^6 = 64$) which is $(0 \equiv 000000, 16 \equiv 010000, 56 \equiv 111000, 40 \equiv 
  101000)$. 
  The white color indicates the dynamics where the largest basin of attraction was found to be that 
  of the dominant limit cycle, the green color stands for all attraction basins having the same 
  size, and the yellow color for the largest attraction basin being that of a non-dominant limit 
  cycle.
  Finally, these results show that:
  \begin{enumerate}
  \item The maximum possible number of different dynamics (147) is reached in this network.
  \item There are no fixed points in these $147$ dynamics. 
    Consequently, they have only limit cycles, whose length belongs to the set $\{2, 3, 4, 6\}$. 
    More specifically, Figure~\ref{fig:BAN-UDdyn} shows that:
    \begin{itemize}
    \item[--] $57$ dynamics of type LC-$(4)$, which means that they admit a limit cycle of length 
      $4$: 
      \begin{compactitem}
      \item $19$ of them converge toward the limit cycle defined by the periodic sequence 
      $(000000, 01000, 111000, 101000)$; 
      \item $19$ converge toward the limit cycle $(000000, 011000, 111000, 100000)$; and 
      \item $19$ converge toward the limit cycle $(010000, 011000, 101000, 100000)$.
      \end{compactitem}
    \item[--] $42$ dynamics of type LC-$(2)$, \emph{i.e.} admitting a limit cycle of length $2$:
      \begin{compactitem} 
      \item $14$ of them converge toward the limit cycle $(000000, 111000)$; 
      \item $14$ converge toward the limit cycle $(010000, 101000)$; and 
      \item $14$ converge toward the limit cycle $(011000, 100000)$.
      \end{compactitem}
    \item[--] $18$ dynamics of type LC-$(2,2)$, \emph{i.e.} admitting $2$ limit cycles of length 
      $2$. 
      Among these dynamics, $6$ have one of their two limit cycles being 
      $(000000, 111000)$, $6$ have one of their two limit cycles being $(010000, 101000)$, and 
      $6$ have one of their two limit cycles being $(011000, 100000)$.
    \item[--] $15$ dynamics of type LC-$(2,6)$, \emph{i.e.} admitting $2$ limit cycles, one of 
      length $2$, one of length $6$. 
      All of them have the same $2$ limit cycles: $(00100, 110000)$ and 
      $(000000, 010000, 011000, 111000, 110000, 100000)$. 
      One of these $15$ dynamics is the parallel one.
    \item[--] $6$ dynamics are of type LC-$(2,2,6)$, \emph{i.e.} admitting $3$ limit cycles, two of 
      length $2$, one of length $6$. 
      All of them have $(000000, 010000, 011000, 111000, 110000, 100000)$ and $(001000, 110000)$ as 
      limit cycles.
    \item[--] $6$ dynamics are of type LC-$(3,4)$, \emph{i.e.} admitting $2$ limit cycles, one of 
      length $3$, one of length $4$. 
      Among these dynamics, $2$ have one of their two limit cycles being 
      $(000000, 010000, 111000, 110000)$, $2$ have one of their two limit cycles being 
      $(000000, 011000, 111000, 100000)$, and $2$ dynamics have one of their two limit cycles being 
      $(010000, 011000, 101\-000, 100000)$.
    \item[--] $3$ dynamics are of type LC-$(2,3)$, \emph{i.e.} admitting $2$ limit cycles, one of 
      length $2$, one of length $3$.
      Among these dynamics, $1$ has the limit cycles $(000000, 111000)$ and 
      $(000001, 111100, 101010)$, $1$ dynamic has the limit cycles $(010000, 101000)$ and 
      $(101010, 010001, 101100)$, and $1$ has the limit cycles $(011000, 100000)$ and 
      $(011100, 101010, 100001)$. 
    \end{itemize}
  \item From Figure~\ref{fig:BAN-UDdyn}, it can be checked that:
    \begin{itemize}
    \item[--] $\{000000, 010000, 011000, 100000, 101000, 111000\}$ is a dominant set of the network.
    \item[--] These $6$ dominant configurations are equally distributed in each type of dynamics. 
      Notably, each of them appears 
      in $38$ of the $57$ dynamics of type LC-$(4)$, 
      in $14$ of the $42$ dynamics of type LC-$(2)$, 
      in $6$ of the $18$ dynamics of type LC-$(2,2)$, 
      in $15$ of the $15$ dynamics of type LC-$(2,6)$, 
      in $6$ of the $6$ dynamics of type LC-$(2,2,6)$, 
      in $4$ of the $6$ dynamics of type LC-$(3,4)$, and 
      in $1$ of the $3$ dynamics of type LC-$(2,3)$. 
      More globally, each of these $6$ dominant configurations appears in $84$ of the $147$ 
      different block-sequential dynamics of the network. 
      Notice that the largest attraction basins are found precisely in the dominant limit cycles. 
      Specifically in $125$ dynamics (those colored in white in~\ref{fig:BAN-UDdyn} which represent 
      $85$\% of the set of possible dynamics).
      Then, in another $15$ dynamics (those colored in yellow which represent $10$\% of the total), 
      the largest attraction basin was found in a non-dominant limit cycle. 
      Finally, in the remaining $7$ dynamics (those colored in green which represent $5$\% of the 
      total), attraction basins of all attractors are of same size.
    \end{itemize}
  \end{enumerate}
\end{enumerate}

In view of the results depicted in Figure~\ref{fig:BAN-UDdyn}, we could say that network $\Psi$ is 
robust to block-sequential update modes. Indeed, regardless of the update mode used, fixed points 
never appear. 
Now, if we consider the parallel dynamics of $\Psi$ which is of type LC-$(2,6)$, it is interesting 
to ask if we can make small modifications to $\Psi$ while maintaining its graph and type of local 
functions in order to obtain another network whose parallel dynamics is also of type LC-$(2,6)$.
A first step in doing this and finding networks like $\Psi$ that eventually have no fixed points is 
considering networks whose interaction graphs contain at least one negative edge, as stated by the 
following proposition.

\begin{prop}\label{prop:neg-edge}
  Let $f$ be a Boolean automata network. 
  Let $g$ be a Boolean automata network whose interaction graph is the same as that of $f$. 
  If a block-sequential dynamics of $g$ has no fixed points, then neither do its other 
  block-sequential dynamics, and at least one of the edges of $g$ is negative.
\end{prop}

\begin{proof}
  Consider $f$ and $g$ two distinct Boolean automata networks as stated just above. 
  Let $\mu$ be a block-sequential update mode such that the dynamics of $g$ with $\mu$ has no fixed 
  point. 
  Since fixed points are invariant regardless of the block-sequential update 
  mode~\cite{Robert1986}, 
  then it is direct that the other block-sequential dynamics do not have any either. 
  Now, suppose that all edges of $g$ are positive. 
  Since the type of local functions are of type \textsc{and/or}, then the all-$0$ and all-$1$ 
  configurations are fixed points of $g$, which is a contradiction.
\end{proof}

Proposition~\ref{prop:neg-edge} establishes a starting condition to eventually find the networks 
having at least one negative edge. However, in the framework of our model $\Psi$, it is easy to do 
an exhaustive search by simulation and show that there are no other like $\Psi$. 
Indeed, without entering into the details, all possible \textsc{and/or} Boolean automata networks 
having an interaction graph which conserves the structure of that $\Psi$ admit parallel dynamics 
which include asymptotically either fixed points or limit cycles of other type.\medskip

\begin{figure}[t!]
  \begin{center}
    \begin{minipage}{.28\textwidth}
      \centerline{\footnotesize $\alpha: B^6 \to B^6$}\medskip
			
			\centerline{\footnotesize
        $\alpha(x) = \left\{\begin{array}{l}
          \alpha_1(x) = x_3\\
          \alpha_2(x) = x_1\\
          \alpha_3(x) = x_2 \lor x_4\\
          \alpha_4(x) = \neg x_3 \land x_6\\
          \alpha_5(x) = x_4\\
          \alpha_6(x) = x_5\\
        \end{array}\right.$}
    \end{minipage}
    \begin{minipage}{.15\textwidth}
      	\centerline{\scalebox{.8}{\begin{tikzpicture}[>=latex,auto]
        \tikzstyle{node} = [draw, circle, thick]
        \node[node](n1) at (-1.5,0) {1};
        \node[node](n2) at (0,0) {2};
        \node[node](n3) at (-.75,-1) {3};
        \node[node](n4) at (-.75,-2.5) {4};
        \node[node](n5) at (0,-3.5) {5};
        \node[node](n6) at (-1.5,-3.5) {6};
        
				\draw[Green4, thick, ->] (n1) edge node {$+$} (n2);
				\draw[Green4, thick, ->] (n2) edge node {$+$} (n3);
				\draw[Green4, thick, ->] (n3) edge node {$+$} (n1);
				\draw[Red3, thick, ->] (n3) edge [bend left=20] node {$-$} (n4);
				\draw[Green4, thick, ->] (n4) edge [bend left=20] node {$+$} (n3);
				\draw[Green4, thick, ->] (n4) edge node {$+$} (n5);
				\draw[Green4, thick, ->] (n5) edge node {$+$} (n6);
				\draw[Green4, thick, ->] (n6) edge node {$+$} (n4);
			\end{tikzpicture}}}
    \end{minipage}
    \hspace*{10mm}
    \begin{minipage}{.28\textwidth}
      \centerline{\footnotesize $\beta: B^6 \to B^6$}\medskip
			
      \centerline{\footnotesize
        $\beta(x) = \left\{\begin{array}{l}
          \beta_1(x) = x_3\\
          \beta_2(x) = \neg x_1\\
          \beta_3(x) = x_2 \lor x_4\\
          \beta_4(x) = \neg x_3 \land x_6\\
          \beta_5(x) = x_4\\
          \beta_6(x) = \neg x_5\\
        \end{array}\right.$
      }      
    \end{minipage}
    \begin{minipage}{.15\textwidth}
      	\centerline{\scalebox{.8}{\begin{tikzpicture}[>=latex,auto]
        \tikzstyle{node} = [draw, circle, thick]
        \node[node](n1) at (-1.5,0) {1};
        \node[node](n2) at (0,0) {2};
        \node[node](n3) at (-.75,-1) {3};
        \node[node](n4) at (-.75,-2.5) {4};
        \node[node](n5) at (0,-3.5) {5};
        \node[node](n6) at (-1.5,-3.5) {6};
        
				\draw[Red3, thick, ->] (n1) edge node {$-$} (n2);
				\draw[Green4, thick, ->] (n2) edge node {$+$} (n3);
				\draw[Green4, thick, ->] (n3) edge node {$+$} (n1);
				\draw[Red3, thick, ->] (n3) edge [bend left=20] node {$-$} (n4);
				\draw[Green4, thick, ->] (n4) edge [bend left=20] node {$+$} (n3);
				\draw[Green4, thick, ->] (n4) edge node {$+$} (n5);
				\draw[Red3, thick, ->] (n5) edge node {$-$} (n6);
				\draw[Green4, thick, ->] (n6) edge node {$+$} (n4);
			\end{tikzpicture}}}
    \end{minipage}
  \end{center}

  \begin{center}
    \begin{minipage}{.28\textwidth}
      \centerline{\footnotesize $\gamma: B^6 \to B^6$}\medskip
			
			\centerline{\footnotesize
        $\gamma(x) = \left\{\begin{array}{l}
          \gamma_1(x) = x_3\\
          \gamma_2(x) = \neg x_1\\
          \gamma_3(x) = x_2 \lor \neg x_4\\
          \gamma_4(x) = x_3 \land x_6\\
          \gamma_5(x) = x_4\\
          \gamma_6(x) = x_5\\
        \end{array}\right.$}
    \end{minipage}
    \begin{minipage}{.15\textwidth}
      	\centerline{\scalebox{.8}{\begin{tikzpicture}[>=latex,auto]
        \tikzstyle{node} = [draw, circle, thick]
        \node[node](n1) at (-1.5,0) {1};
        \node[node](n2) at (0,0) {2};
        \node[node](n3) at (-.75,-1) {3};
        \node[node](n4) at (-.75,-2.5) {4};
        \node[node](n5) at (0,-3.5) {5};
        \node[node](n6) at (-1.5,-3.5) {6};
        
				\draw[Red3, thick, ->] (n1) edge node {$-$} (n2);
				\draw[Green4, thick, ->] (n2) edge node {$+$} (n3);
				\draw[Green4, thick, ->] (n3) edge node {$+$} (n1);
				\draw[Green4, thick, ->] (n3) edge [bend left=20] node {$+$} (n4);
				\draw[Red3, thick, ->] (n4) edge [bend left=20] node {$-$} (n3);
				\draw[Green4, thick, ->] (n4) edge node {$+$} (n5);
				\draw[Green4, thick, ->] (n5) edge node {$+$} (n6);
				\draw[Green4, thick, ->] (n6) edge node {$+$} (n4);
			\end{tikzpicture}}}
    \end{minipage}
    \hspace*{10mm}
    \begin{minipage}{.28\textwidth}
      \centerline{\footnotesize $\delta: B^6 \to B^6$}\medskip
			
      \centerline{\footnotesize
        $\delta(x) = \left\{\begin{array}{l}
          \delta_1(x) = x_3\\
          \delta_2(x) = x_1\\
          \delta_3(x) = x_2 \lor \neg x_4\\
          \delta_4(x) = x_3 \land x_6\\
          \delta_5(x) = x_4\\
          \delta_6(x) = x_5\\
        \end{array}\right.$
      }      
    \end{minipage}
    \begin{minipage}{.15\textwidth}
      	\centerline{\scalebox{.8}{\begin{tikzpicture}[>=latex,auto]
        \tikzstyle{node} = [draw, circle, thick]
        \node[node](n1) at (-1.5,0) {1};
        \node[node](n2) at (0,0) {2};
        \node[node](n3) at (-.75,-1) {3};
        \node[node](n4) at (-.75,-2.5) {4};
        \node[node](n5) at (0,-3.5) {5};
        \node[node](n6) at (-1.5,-3.5) {6};
        
				\draw[Green4, thick, ->] (n1) edge node {$+$} (n2);
				\draw[Green4, thick, ->] (n2) edge node {$+$} (n3);
				\draw[Green4, thick, ->] (n3) edge node {$+$} (n1);
				\draw[Green4, thick, ->] (n3) edge [bend left=20] node {$+$} (n4);
				\draw[Red3, thick, ->] (n4) edge [bend left=20] node {$-$} (n3);
				\draw[Green4, thick, ->] (n4) edge node {$+$} (n5);
				\draw[Green4, thick, ->] (n5) edge node {$+$} (n6);
				\draw[Green4, thick, ->] (n6) edge node {$+$} (n4);
			\end{tikzpicture}}}
    \end{minipage}  
  \end{center}

  \begin{center}
    \begin{minipage}{.28\textwidth}
      \centerline{\footnotesize $\zeta: B^6 \to B^6$}\medskip
			
			\centerline{\footnotesize
        $\zeta(x) = \left\{\begin{array}{l}
          \zeta_1(x) = x_3\\
          \zeta_2(x) = \neg x_1\\
          \zeta_3(x) = x_2 \lor \neg x_4\\
          \zeta_4(x) = x_3 \land x_6\\
          \zeta_5(x) = x_4\\
          \zeta_6(x) = \neg x_5\\
        \end{array}\right.$}
    \end{minipage}
    \begin{minipage}{.15\textwidth}
      	\centerline{\scalebox{.8}{\begin{tikzpicture}[>=latex,auto]
        \tikzstyle{node} = [draw, circle, thick]
        \node[node](n1) at (-1.5,0) {1};
        \node[node](n2) at (0,0) {2};
        \node[node](n3) at (-.75,-1) {3};
        \node[node](n4) at (-.75,-2.5) {4};
        \node[node](n5) at (0,-3.5) {5};
        \node[node](n6) at (-1.5,-3.5) {6};
        
				\draw[Red3, thick, ->] (n1) edge node {$-$} (n2);
				\draw[Green4, thick, ->] (n2) edge node {$+$} (n3);
				\draw[Green4, thick, ->] (n3) edge node {$+$} (n1);
				\draw[Green4, thick, ->] (n3) edge [bend left=20] node {$+$} (n4);
				\draw[Red3, thick, ->] (n4) edge [bend left=20] node {$-$} (n3);
				\draw[Green4, thick, ->] (n4) edge node {$+$} (n5);
				\draw[Red3, thick, ->] (n5) edge node {$-$} (n6);
				\draw[Green4, thick, ->] (n6) edge node {$+$} (n4);
			\end{tikzpicture}}}
    \end{minipage}
    \hspace*{10mm}
    \begin{minipage}{.28\textwidth}
      \centerline{\footnotesize $\eta: B^6 \to B^6$}\medskip
			
      \centerline{\footnotesize
        $\eta(x) = \left\{\begin{array}{l}
          \eta_1(x) = x_3\\
          \eta_2(x) = \neg x_1\\
          \eta_3(x) = x_2 \lor x_4\\
          \eta_4(x) = x_3 \land x_6\\
          \eta_5(x) = x_4\\
          \eta_6(x) = x_5\\
        \end{array}\right.$
      }      
    \end{minipage}
    \begin{minipage}{.15\textwidth}
      	\centerline{\scalebox{.8}{\begin{tikzpicture}[>=latex,auto]
        \tikzstyle{node} = [draw, circle, thick]
        \node[node](n1) at (-1.5,0) {1};
        \node[node](n2) at (0,0) {2};
        \node[node](n3) at (-.75,-1) {3};
        \node[node](n4) at (-.75,-2.5) {4};
        \node[node](n5) at (0,-3.5) {5};
        \node[node](n6) at (-1.5,-3.5) {6};
        
				\draw[Red3, thick, ->] (n1) edge node {$-$} (n2);
				\draw[Green4, thick, ->] (n2) edge node {$+$} (n3);
				\draw[Green4, thick, ->] (n3) edge node {$+$} (n1);
				\draw[Green4, thick, ->] (n3) edge [bend left=20] node {$+$} (n4);
				\draw[Green4, thick, ->] (n4) edge [bend left=20] node {$+$} (n3);
				\draw[Green4, thick, ->] (n4) edge node {$+$} (n5);
				\draw[Green4, thick, ->] (n5) edge node {$+$} (n6);
				\draw[Green4, thick, ->] (n6) edge node {$+$} (n4);
			\end{tikzpicture}}}
    \end{minipage}  
  \end{center}
  \caption{Six \textsc{and/or} Boolean automata networks with their associated interactions which 
    conserve both the type of local functions of $\Psi$ and the underlying structure of that of 
    $\Psi$.}
  \label{fig:6BANs}
\end{figure}
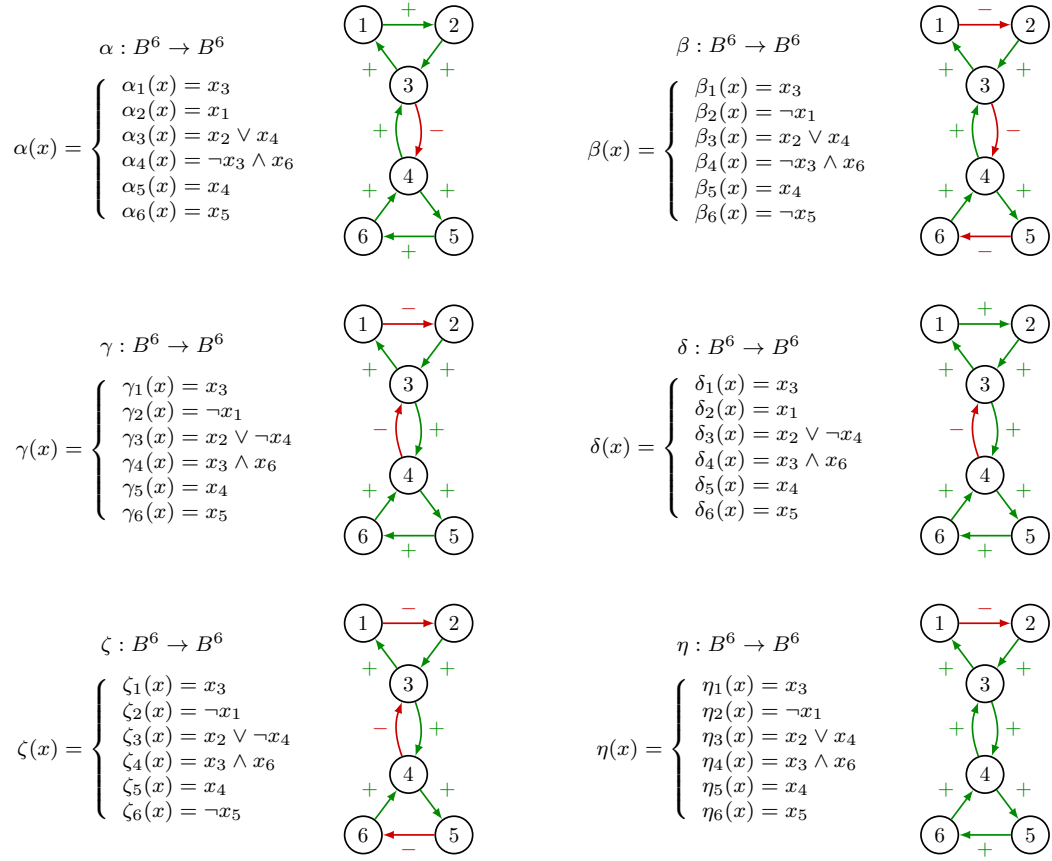

As an illustration, Figure~\ref{fig:6BANs} depicts six Boolean automata networks together with their 
interaction graphs which are small modifications of $\Psi$. 
If we compute all the block-sequential dynamics of each of these networks, we can observe the 
following differences with those of $\Psi$:
\begin{itemize}
\item[--] The parallel dynamics of $\alpha$ and $\delta$ includes two fixed points and there exist 
  other dynamics with no limit cycles; 
\item[--] The parallel dynamics of $\beta$ and $\zeta$ has no fixed point but has a limit cycle of 
  length $5$;
\item[--] The parallel dynamics of $\gamma$ and $\eta$ includes one fixed point and there exist 
  other dynamics with no limit cycles.
\end{itemize}

\subsection{Discussion around a specific intricate update mode}\label{sec:results:discussion}

We have already seen from results of Section~\ref{sec:results:robustness} that Boolean 
automata network $\Psi$ related to hereditary angioedema is robust because, regardless of the 
block-sequential update mode, fixed points never appear, which is coherent with the empiric 
observation that most of genes involved the disease have a periodic expression due in part to the 
chromatin clock and circadian rhythms~\cite{Zhang2020}.\medskip

Some genes are updated systematically together under the control of the chromatin clock.
In order to achieve a systematic plan of numerical experience, we shall compare the block-sequential 
update modes related to $\Psi$ studied in Section~\ref{sec:results:robustness} to an intricate 
update mode, for which we consider that automata $3$ (modeling SP1) and $4$ (modeling F12) tends to 
actively express their protein. 
This intricate update mode is a specific block-parallel update mode which has been inspired by 
F.~Robert, who defined the notion of interactions between two Boolean automata networks of size $2$ 
as follows. 
Consider two Boolean automata networks of size $2$ whose global functions are $f$ and $g$. 
Interactions between $f$ and $g$ can be of four types~\cite{Demongeot2022}:
\begin{itemize}
\item[--] Sequential dependence: two networks $f$ and $g$ of size $2$ are said to be 
  sequential-dependent when
  \begin{multline*}
    f_1(x_1 x_2) = f_0(g_0(x_1 x_2) x_2), g_1(x_1 x_2) = g_0(x_1 f_1(x_1 x_2)), \dots,\\
    f_i(x_1 x_2) = f_{i-1}(g_{i-1}(x_1 x_2) x_2), g_i(x_1 x_2) = g_{i-1}(x_1 f_i(x_1 x_2))\text{,}
  \end{multline*}
  or conversely when
  \begin{multline*}
    g_1(x_1 x_2) = g_0(f_0(x_1 x_2) x_2), f_1(x_1 x_2) = f_0(x_1 g_1(x_1 x_2)), \dots,\\
    g_i(x_1 x_2) = g_{i-1}(f_{i-1}(x_1 x_2) x_2), f_i(x_1 x_2) = f_{i-1}(x_1 g_i(x_1 x_2))\text{.}
  \end{multline*}
\item[--] Parallel dependence: two networks $f$ and $g$ of size $2$ are said to be 
  parallel-dependent when
  \begin{multline*}
    f_1(x_1 x_2) = f_0(g_0(x_1 x_2) x_2), g_1(x_1 x_2) = g_0(x_1 f_0(x_1 x_2)), \dots,\\ 
    f_i(x_1 x_2) = f_{i-1}(g_{i-1}(x_1 x_2) x_2), g_i(x_1 x_2) = 
    g_{i-1}(x_1 f_{i-1}(x_1 x_2))\text{,}
  \end{multline*}
  or conversely when 
  \begin{multline*}
    g_1(x_1 x_2) = g_0(f_0(x_1 x_2) x_2), f_1(x_1 x_2) = f_0(x_1 g_0(x_1 x_2)), \dots,\\ 
    g_i(x_1 x_2) = g_{i-1}(f_{i-1}(x_1 x_2) x_2), f_i(x_1 x_2) = 
    f_{i-1}(x_1 g_{i-1}(x_1 x_2))\text{.}
  \end{multline*}
\item[--] Sequential opposition: two networks $f$ and $g$ of size $2$ are said to be 
  sequential-opposed when
  \begin{multline*}
    f_1(x_1 x_2) = f_0(\neg g_0(x_1 x_2) x_2), g_1(x_1 x_2) = g_0(x_1 \neg f_1(x_1 x_2)), \dots,\\ 
    f_i(x_1 x_2) = f_{i-1}(\neg g_{i-1}(x_1 x_2) x_2), g_i(x_1 x_2) = 
    g_{i-1}(x_1 \neg f_i(x_1 x_2))\text{.}
  \end{multline*}
  or conversely when
  \begin{multline*}
    g_1(x_1 x_2) = g_0(\neg f_0(x_1 x_2) x_2), f_1(x_1 x_2) = f_0(x_1 \neg g_1(x_1 x_2)), \dots,\\ 
    g_i(x_1 x_2) = g_{i-1}(\neg f_{i-1}(x_1 x_2) x_2), f_i(x_1 x_2) = 
    f_{i-1}(x_1 \neg g_i(x_1 x_2))\text{.}
  \end{multline*}
\item[--] Parallel opposition: two networks $f$ and $g$ of size $2$ are said to be parallel-opposed 
  when:
  \begin{multline*}
    f_1(x_1 x_2) = f_0(\neg g_0(x_1 x_2) x_2), g_1(x_1 x_2) = g_0(x_1 \neg f_0(x_1 x_2)), \dots,\\ 
    f_i(x_1 x_2) = f_{i-1}(\neg g_{i-1}(x_1 x_2) x_2), g_i(x_1 x_2) = 
    g_{i-1}(x_1 \neg f_{i-1}(x_1 x_2))\text{,}
  \end{multline*}
  or conversely when
  \begin{multline*}
    g_1(x_1 x_2) = g_0(\neg f_0(x_1 x_2) x_2), f_1(x_1 x_2) = f_0(x_1 \neg g_0(x_1 x_2)), \dots,\\ 
    g_i(x_1 x_2) = g_{i-1}(\neg f_{i-1}(x_1 x_2) x_2), f_i(x_1 x_2) = 
    f_{i-1}(x_1 \neg g_{i-1}(x_1 x_2))\text{.}
  \end{multline*}
\end{itemize}

We can also write for the sequential dependence between automata networks of size $n$: 
\begin{multline*}
  g_0 = g, f_0 = f, g_1 = g_0(f_{0, I} K), f_1 = f_0(I g_{1, K}), \dots,\\
  g_i = g_{i-1}(f_{i-1, I} K), f_i = f_{i-1}(I g_{i-1, K}), \dots,\\
  g_n = g_{n-1}(f_{n-1, I} K), f_n = f_{n-1}(I g_{n-1, K})\text{,}
\end{multline*} 
where $I$ and $K$ $\subseteq \{1, \dots, n\}$ and $I$ (resp. $K$) is the identity on 
$x_{I\setminus K} = \{x_i\}_{i \in I\setminus K}$ (resp. $x_{K\setminus I} = 
\{x_j\}_{j K\setminus I}$, and $f_{0,I}$ (resp. $g_{1,K}$) is the projection of $f_0$ on $I\setminus 
K$ (resp. $g_1$ on $K\setminus I$).\medskip

In our case, $n = 6$ and $f$ and $g$ are defined by the logic equations of the 
Figure~\ref{fig:BAN-GRN}. 
Complex biological networks fall within the general framework above, 
provided that more information is available on the role of the chromatin clock 
which regulates the timing of updating of their components.
This constitutes an additional degree of complexity because certain histones involved in this 
chromatin regulation have transcription factors involved in the networks whose expression is 
controlled by these histones.\medskip
 
F. Robert studied systematically the corresponding dynamics and proved the following proposition in 
the case of a network of size $2$, when $I \cap K \neq \emptyset$.

\begin{prop}\label{prop:robert}\ 
  \begin{enumerate}
  \item Sequential and parallel dependence (resp. opposition) rules give same fixed points.
  \item The update mode defined by the sequential (resp. parallel) opposition between $\neg f$ and 
    $\neg g$ gives the complementary dynamics to that given by the update mode defined by the 
    sequential (resp. parallel) dependence between $f$ and $g$.
  \end{enumerate}
\end{prop}

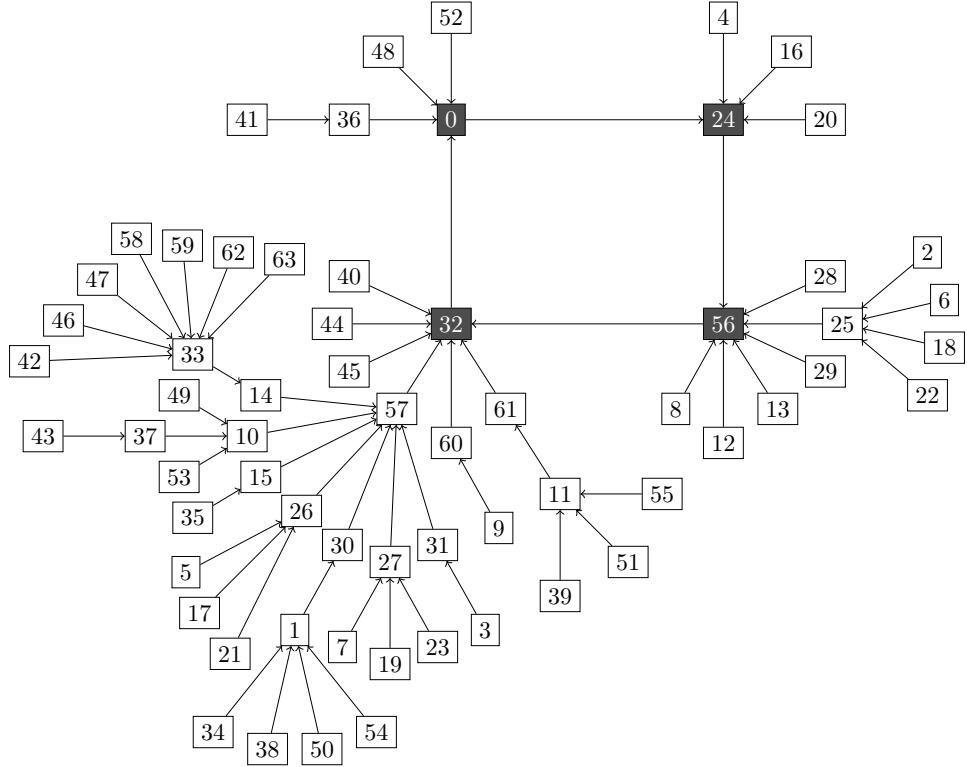
\begin{figure}[t!]
  \begin{center}
    \scalebox{.9}{\begin{tikzpicture}[>=to,auto]
      \tikzstyle{conf} = [rectangle, draw]
      \tikzstyle{pf} = [rectangle, draw, fill=black!10]
      \tikzstyle{lc} = [rectangle, draw, fill=black!70]
      \node[lc](n0) at (0,1) {\textcolor{white}{$0$}};
      \node[conf](n1) at (-2.3,-6.5) {$1$};
      \node[conf](n2) at (7,-.95) {$2$};
      \node[conf](n3) at (.5,-6.5) {$3$};
      \node[conf](n4) at (4,2.5) {$4$};
      \node[conf](n5) at (-3.9,-5.65) {$5$};
      \node[conf](n6) at (7.25,-1.65) {$6$};
      \node[conf](n7) at (-1.6,-6.75) {$7$};
      \node[conf](n8) at (3.3,-3.25) {$8$};
      \node[conf](n9) at (.7,-5) {$9$};
      \node[conf](n10) at (-3,-3.65) {$10$};
      \node[conf](n11) at (1.6,-4.5) {$11$};
      \node[conf](n12) at (4,-3.75) {$12$};
      \node[conf](n13) at (4.8,-3.25) {$13$};
      \node[conf](n14) at (-2.8,-3.05) {$14$};
      \node[conf](n15) at (-2.8,-4.25) {$15$};
      \node[conf](n16) at (5,2) {$16$};
      \node[conf](n17) at (-3.7,-6.25) {$17$};
      \node[conf](n18) at (7.25,-2.35) {$18$};
      \node[conf](n19) at (-.9,-7) {$19$};
      \node[conf](n20) at (5.5,1) {$20$};
      \node[conf](n21) at (-3.25,-6.85) {$21$};
      \node[conf](n22) at (7,-3.05) {$22$};
      \node[conf](n23) at (-.2,-6.75) {$23$};
      \node[lc](n24) at (4,1) {\textcolor{white}{$24$}};
      \node[conf](n25) at (5.75,-2) {$25$};
      \node[conf](n26) at (-2.2,-4.75) {$26$};
      \node[conf](n27) at (-.9,-5.5) {$27$};
      \node[conf](n28) at (5.5,-1.3) {$28$};
      \node[conf](n29) at (5.5,-2.7) {$29$};
      \node[conf](n30) at (-1.6,-5.25) {$30$};
      \node[conf](n31) at (-.2,-5.25) {$31$};
      \node[lc](n32) at (0,-2) {\textcolor{white}{$32$}};
      \node[conf](n33) at (-3.8,-2.45) {$33$};
      \node[conf](n34) at (-3.5,-8) {$34$};
      \node[conf](n35) at (-3.8,-4.85) {$35$};
      \node[conf](n36) at (-1.5,1) {$36$};
      \node[conf](n37) at (-4.5,-3.65) {$37$};
      \node[conf](n38) at (-2.7,-8.25) {$38$};
      \node[conf](n39) at (1.6,-6) {$39$};
      \node[conf](n40) at (-1.5,-1.3) {$40$};
      \node[conf](n41) at (-3,1) {$41$};
      \node[conf](n42) at (-6.2,-2.55) {$42$};
      \node[conf](n43) at (-6,-3.65) {$43$};
      \node[conf](n44) at (-1.75,-2) {$44$};
      \node[conf](n45) at (-1.5,-2.7) {$45$};
      \node[conf](n46) at (-5.7,-1.95) {$46$};
      \node[conf](n47) at (-5.2,-1.35) {$47$};
      \node[conf](n48) at (-1,2) {$48$};
      \node[conf](n49) at (-4,-3.05) {$49$};
      \node[conf](n50) at (-1.9,-8.25) {$50$};
      \node[conf](n51) at (2.6,-5.5) {$51$};
      \node[conf](n52) at (0,2.5) {$52$};
      \node[conf](n53) at (-4,-4.25) {$53$};
      \node[conf](n54) at (-1.1,-8) {$54$};
      \node[conf](n55) at (3.1,-4.5) {$55$};
      \node[lc](n56) at (4,-2) {\textcolor{white}{$56$}};
      \node[conf](n57) at (-.8,-3.25) {$57$};
      \node[conf](n58) at (-4.7,-.75) {$58$};
      \node[conf](n59) at (-3.95,-.85) {$59$};
      \node[conf](n60) at (0,-3.75) {$60$};
      \node[conf](n61) at (.8,-3.25) {$61$};
      \node[conf](n62) at (-3.2,-.95) {$62$};
      \node[conf](n63) at (-2.45,-1.05) {$63$};
      
      \draw[->] (n0) edge (n24);
      \draw[->] (n24) edge (n56);
      \draw[->] (n56) edge (n32);
      \draw[->] (n32) edge (n0);
      \draw[->] (n36) edge (n0);
      \draw[->] (n48) edge (n0);
      \draw[->] (n52) edge (n0);
      \draw[->] (n41) edge (n36);
      \draw[->] (n4) edge (n24);
      \draw[->] (n16) edge (n24);
      \draw[->] (n20) edge (n24);
      \draw[->] (n40) edge (n32);
      \draw[->] (n44) edge (n32);
      \draw[->] (n45) edge (n32);
      \draw[->] (n57) edge (n32);
      \draw[->] (n60) edge (n32);
      \draw[->] (n61) edge (n32);
      \draw[->] (n9) edge (n60);
      \draw[->] (n11) edge (n61);
      \draw[->] (n26) edge (n57);
      \draw[->] (n10) edge (n57);
      \draw[->] (n14) edge (n57);
      \draw[->] (n15) edge (n57);
      \draw[->] (n27) edge (n57);
      \draw[->] (n30) edge (n57);
      \draw[->] (n31) edge (n57);
      \draw[->] (n37) edge (n10);
      \draw[->] (n49) edge (n10);
      \draw[->] (n53) edge (n10);
      \draw[->] (n33) edge (n14);
      \draw[->] (n35) edge (n15);
      \draw[->] (n7) edge (n27);
      \draw[->] (n19) edge (n27);
      \draw[->] (n23) edge (n27);
      \draw[->] (n1) edge (n30);
      \draw[->] (n3) edge (n31);
      \draw[->] (n17) edge (n26);
      \draw[->] (n5) edge (n26);
      \draw[->] (n21) edge (n26);
      \draw[->] (n8) edge (n56);
      \draw[->] (n12) edge (n56);
      \draw[->] (n13) edge (n56);
      \draw[->] (n25) edge (n56);
      \draw[->] (n28) edge (n56);
      \draw[->] (n29) edge (n56);
      \draw[->] (n2) edge (n25);
      \draw[->] (n6) edge (n25);
      \draw[->] (n18) edge (n25);
      \draw[->] (n22) edge (n25);
      \draw[->] (n34) edge (n1);
      \draw[->] (n38) edge (n1);
      \draw[->] (n50) edge (n1);
      \draw[->] (n54) edge (n1);
      \draw[->] (n39) edge (n11);
      \draw[->] (n51) edge (n11);
      \draw[->] (n55) edge (n11);
      \draw[->] (n43) edge (n37);
      \draw[->] (n42) edge (n33);
      \draw[->] (n46) edge (n33);
      \draw[->] (n47) edge (n33);
      \draw[->] (n58) edge (n33);
      \draw[->] (n59) edge (n33);
      \draw[->] (n62) edge (n33);
      \draw[->] (n63) edge (n33);
		\end{tikzpicture}}
  \end{center}
  \caption{Dynamics of BAN$_6$ $f$ presented in Figure~\ref{fig:BAN-GRN} according to the intricate 
    block-parallel update mode $(\{1, 2, 3, 4\}, \{3, 4, 5, 6\})$ which exhibits a unique limit 
    cycle composed of dark configurations. The $64$ configurations are represented thanks to their 
    ten-base encoding for the sake of concision.}
  \label{fig:BAN_BPdyn}
\end{figure}

We can remark on Figure~\ref{fig:BAN_BPdyn} that there is a unique limit cycle $\text{LC}$, which is 
of length $4$ for the intricate updating mode $(\{1, 2, 3, 4\}, \{3, 4, 5, 6\})$ which implements 
the ideas behind the sequential dependence defined above, where 
$LC = (100000, 000000, 011000, 111000)$. 
This limit cycle corresponds to the most frequent asymptotic behavior of the non-intricate 
block-sequential dynamics (see Figure~\ref{fig:BAN-UDdyn}), which reinforces the robust character of 
the genetic network, considered as the core of the regulation involved in the hereditary angioedema.

If there is a change in the expression of the genes involved in this disease, it will not be due to 
a change in the clock (for example, the chromatin clock or the circadian clock that controls it), 
but rather, which confirms the currently accepted explanation, to an endogenous hormonal influence 
(estrogen impregnation) or to exogenous elements triggering attacks of a physical nature (influence 
of cold or exercise), dietary (allergy) or pathological (stress; surgical or drug therapy; common 
illnesses such as colds and flu; insect bites, etc.). 

\section{Conclusion}
\label{sec:conclusion}

Before we really conclude this article, let us give precisions about the relevance of using Boolean 
modeling in this kind of framework. 
Indeed, one might question whether the binary representation of genes, which disregards continuous 
expression levels and potential quantitative effects, is a limitation of our approach. 
Actually, we argue that Boolean networks are a powerful and general modeling framework, far from 
being a limiting choice. 
Indeed, they can represent any real system composed of interacting entities whose states are not 
directly observable, only the completed changes between states matter. 
This makes them applicable to a very wide range of systems.
Moreover, from a modeling perspective, the choice of formalism depends on how the variable of 
interest is interpreted. 
Using gene expression as an example, one could choose to use Boolean, discrete or continuous (or 
hybrid) modelling depending on whether we represent gene on/off expression, several interaction 
levels genes can have on one another, or the concentration of proteins they produce.
All these approaches can ultimately be handled within the Boolean framework, either directly or 
through encoding and approximation, so that this flexibility makes Boolean models not inherently 
reductive.
Also, biologists naturally express regulatory mechanisms in logical terms ("if... then...", 
presence/absence of expression or silencing of a gene, etc.), which further supports the relevance 
of Boolean modelling for genetic regulation networks.
Finally, we really think that Boolean models are well-suited for qualitative analyses where state 
changes matter more than the states themselves.
Furthermore, from a theoretical standpoint, Boolean networks were proven in~\cite{Goles1990} to be 
universal models of computation, meaning they can compute any computable function, making them 
mathematically and computationally far from limited.\medskip

That being said, in this article, we have also introduced a new variety of periodic update modes, 
the intricate update modes, which generalizes that of block-parallel update modes by inclusion. 
Indeed, from the mathematical standpoint, block-parallel update modes are modes with local 
repetitions of updates over their period which are derived from their specific characterization as 
partitioned orders whereas intricate modes admit local repetitions of updates with no specific 
structural constraint.\medskip

Moreover, from the biological modeling standpoint, basing ourselves on the wish to make advances on 
the understanding of the genetic regulation governing hereditary angioedema, we developed a 
simplified version of a known genetic regulatory network for this disease, which conserves only 
the major players in regulation of gene expression.

From this network and its underlying Boolean network model, we emphasized the strong 
robustness of our model with respect to block-sequential update, by showing that none of its 
asymptotic behaviors are fixed points which is coherent with empiric observation that the expression 
of most of the genes involved in hereditary angioedema follows periodical paces.

It stays important to understand that such a study of the robustness of a model of genetic 
regulation network is intrinsically complex and raises scalability issues, even by considering 
``only'' block-sequential update modes.
The first issue is combinatorial: the number of possible block-sequential update modes grows 
exponentially (it follows the Fubini number sequence), making a naive approach consisting in 
computing and comparing all associated dynamical systems completely impractical as soon as the 
network exceeds around ten automata.
The proposed solution relies on update digraphs, objects directly built from the interaction graph 
of the network. 
Theorem~\ref{thm:arac2009} shows that, given a network, two block-sequential update modes sharing 
the same update digraph produce identical dynamical systems. 
Therefore, counting distinct dynamical systems reduces to counting these digraphs, which can be 
tractable in practice for certain Boolean networks.
The remaining limitation is that as soon as one wishes to analyse in detail the properties of the 
obtained dynamical systems, rather than merely counting them, complexity becomes exponential again. 
Consequently, exhaustive analyses such as the one performed in this article are therefore only 
feasible for small networks (of size less than 30 automata approximately).\medskip

Then, we developed an intricate update mode, which is an alternating block-parallel update mode, a
family of update modes introduced recently in~\cite{Demongeot2020} in the framework of biological 
modeling~\cite{Tomassone1993} and highlighted to be able to model the action of Zeitgebers. 
Here, the intricate update mode is particular in the sense that it is structurally defined such that 
a core subnetwork of two automata (or genes) is constantly updated and acts on two other subnetworks 
which alternate their updatings. 
More precisely, the automata of the model are thus updated according to distinct schedules, 
some being updated at each iteration and others being updated at lower frequencies (every other 
time, in the chosen example). 
We highlighted that this intricate update mode captures the main block-sequential limit cycle, 
\emph{i.e.} the one which is most often reached with block-sequential update modes and which admits 
an attraction basin as big as possible, namely composed of all the configurations.\medskip

In the application to Boolean networks of genetic regulation, in the absence of precise information 
on the mode of updating the states of the nodes of the network (i.e. the expression of the genes 
involved), it is necessary to examine the consequences of a change of mode on the dynamics of the 
network (possible change in the number, nature and size of the basins of attraction of its 
attractors). 
If the characteristics of the attractors are invariant in a change of update mode, the network is 
said to be robust. 
In that case, the choice of a dynamics rather an other one is less crucial for the practical 
consequences of the discrete modeling of the expression of the genes studied, in particular to 
explain the genesis of a pathology of familial origin involving a set of well-identified interacting 
genes, which is the case in the disease taken as an example, namely the hereditary 
angioedema.

\paragraph{Acknowledgments}

The authors acknowledge C. Drouet for helpul discussion about hereditary angioedema.
This work was partially funded by 
the ANR-24-CE48-7504 ``ALARICE'' project (S.S.), 
the HORIZON-MSCA-2022-SE-01 101131549 ``ACANCOS'' project (E.G., S.S.), 
the STIC AmSud 22-STIC-02 ``CAMA'' project (E.G., M.M.-M., S.S.), 
the FONDECYT/ANID Regular 1250984 project (E.G., M.M.-M.) and 
the ANID-MILENIO-NCN2024 103 project (E.G.).

%


\begin{thebibliography}{10}

\bibitem{Aracena2013a}
J.~Aracena, J.~Demongeot, {\'E}.~Fanchon, and M.~Montalva.
\newblock {On the number of different dynamics in Boolean networks with
  deterministic update schedules}.
\newblock {\em Mathematical Biosciences}, 242:188--194, 2013.

\bibitem{Aracena2013b}
J.~Aracena, J.~Demongeot, {\'E}.~Fanchon, and M.~Montalva.
\newblock {On the number of update digraphs and its relation with the feedback
  arc sets and tournaments}.
\newblock {\em Discrete Applied Mathematics}, 161:1345--1355, 2013.

\bibitem{Aracena2004}
J.~Aracena, J.~Demongeot, and E.~Goles.
\newblock {Positive and negative circuits in discrete neural networks}.
\newblock {\em IEEE Transactions on Neural Networks}, 15:77--83, 2004.

\bibitem{Aracena2011}
J~Aracena, {\'E}.~Fanchon, M.~Montalva, and M.~Noual.
\newblock {Combinatorics on update digraphs in Boolean networks}.
\newblock {\em Discrete Applied Mathematics}, 159:401--409, 2011.

\bibitem{Aracena2009}
J.~Aracena, E.~Goles, A.~Moreira, and L.~Salinas.
\newblock {On the robustness of update schedules in Boolean networks}.
\newblock {\em Biosystems}, 97:1--8, 2009.

\bibitem{Bhatia1967}
N.~P. Bhatia and G.~P. Szeg{\"o}.
\newblock {\em {Dynamical Systems: Stability Theory and Applications}},
  volume~35 of {\em Lecture Notes in Mathematics}.
\newblock Springer, Berlin, 1967.

\bibitem{Birkhoff1927}
G.~D. Birkhoff.
\newblock {\em {Dynamical Systems}}, volume~9 of {\em AMS Colloquium
  Publications}.
\newblock American Mathematical Society, New York, 1927.

\bibitem{Bowen1975}
R.~Bowen.
\newblock {$\omega$-limit sets for axiom A diffeomorphisms}.
\newblock {\em Journal of Differential Equations}, 18:333--339, 1975.

\bibitem{Conley1978}
C.~Conley.
\newblock {\em {Isolated invariant sets and the Morse index}}, volume~38 of
  {\em Regional Conference Series in Mathematics}.
\newblock American Mathematical Society, Providence, 1978.

\bibitem{Cormen2001}
T.~H. Cormen, C.~E. Leiserson, R.~L. Rivest, and C.~Stein.
\newblock {\em Introduction to Algorithms}.
\newblock MIT Press and McGraw-Hill, Cambridge, 2nd edition, 2001.

\bibitem{Cosnard1985}
M.~Cosnard and J.~Demongeot.
\newblock {Attracteurs\:: une approche d{\'e}terministe}.
\newblock {\em Comptes rendus de l'Acad{\'e}mie des sciences. S{\'e}rie 1,
  Math{\'e}matique}, 300:551--556, 1985.

\bibitem{Demongeot2022}
J.~Demongeot.
\newblock {\em {Seven things I know about them}}, volume~42 of {\em Emergence,
  Complexity and Computation}.
\newblock Springer, Cham, 2022.

\bibitem{Demongeot2008}
J.~Demongeot, A.~Elena, and S.~Sen{\'e}.
\newblock {Robustness in regulatory networks: a multi-disciplinary approach}.
\newblock {\em Acta Biotheoretica}, 56:27--49, 2008.

\bibitem{Demongeot2020}
J.~Demongeot and S.~Sen{\'e}.
\newblock {About block-parallel Boolean networks: a position paper}.
\newblock {\em Natural Computing}, 19:5--13, 2020.

\bibitem{Donoso2025}
I.~Donoso-Leiva, E.~Goles, M.~R{\'i}os-Wilson, and S.~Sen{\'e}.
\newblock {Impact of (a)synchronism on ECA: Towards a new classification}.
\newblock {\em Chaos, Solitons \& Fractals}, 199:116601, 2025.

\bibitem{Garrido1983}
L.~Garrido and C.~Sim{\'o}.
\newblock {Some ideas about strange attractors}.
\newblock In {\em Dynamical Systems and Chaos}, volume 179 of {\em Lecture
  Notes in Physics}, pages 1--28, Berlin, 1983. Springer.

\bibitem{Goles1990}
E.~Goles and S.~Mart{\'i}nez.
\newblock {\em {Neural and Automata Networks: Dynamical Behavior and
  Applications}}.
\newblock Mathematics and Its Application. Kluwer Academic Publishers, 1990.

\bibitem{Goles2018}
E.~Goles, M.~Montalva-Medel, S.~MacLean, and H.~S. Mortveit.
\newblock {Block invariance in a family of elementary cellular automata}.
\newblock {\em Journal of Cellular Automata}, 13:15--32, 2018.

\bibitem{Goles2015}
E.~Goles, M.~Montalva-Medel, H.~S. Mortveit, and S.~Ram{\'i}rez-Flandes.
\newblock {Block invariance in elementary cellular automata}.
\newblock {\em Journal of Cellular Automata}, 10:119--135, 2015.

\bibitem{Goles1980}
E.~Goles and J.~Olivos.
\newblock {Periodic behaviour of generalized threshold functions}.
\newblock {\em Discrete Mathematics}, 30:187--189, 1980.

\bibitem{Guckenheimer1983}
J.~Guckenheimer and P.~Holmes.
\newblock {\em {Nonlinear Oscillations, Dynamical Systems, and Bifurcations of
  Vector Fields}}, volume~42 of {\em Applied Mathematical Sciences}.
\newblock Springer, New York, 1983.

\bibitem{Harvey1997}
I.~Harvey and T.~Bossomaier.
\newblock {Time out of joint, attractors in asynchronous random Boolean
  networks}.
\newblock In {\em ECAL 1997: Fourth European Conference on Artificial Life},
  pages 67--75, Cambridge, 1997. MIT Press.

\bibitem{Hopfield1982}
J.~J. Hopfield.
\newblock {Neural networks and physical systems with emergent collective
  computational abilities}.
\newblock {\em Proceedings of the National Academy od Sciences of the United
  States of America}, 79:2554--2558, 1982.

\bibitem{Kauffman1969}
S.~A. Kauffman.
\newblock {Metabolic stability and epigenesis in randomly constructed genetic
  nets}.
\newblock {\em Journal of Theoretical Biology}, 22:437--467, 1969.

\bibitem{Kleene1951}
S.~C. Kleene.
\newblock {\em {Representation of events in nerve nets and finite automata}}.
\newblock Princeton University Press, Princeton, 1951.

\bibitem{Buffon1749}
G.-L. Leclerc~de Buffon.
\newblock {\em {Histoire naturelle, g{\'e}n{\'e}rale et particuli{\`e}re, avec
  la description du Cabinet du Roy}}.
\newblock Imprimerie Royale, Paris, 1749.

\bibitem{Mayo1989}
C.~F. Mayo and C.~E. Roberts.
\newblock {Sequential, parallel and vector solution of ordinary differential
  equations on a hypercube}.
\newblock {\em Computers \& Mathematics with Applications}, 18:797--808, 1989.

\bibitem{McCulloch1943}
W.~S. McCulloch and W.~Pitts.
\newblock {A logical calculus of the ideas immanent in nervous activity}.
\newblock {\em The Bulletin of Mathematical Biophysics}, 5:115--133, 1943.

\bibitem{Pauleve2022}
L.~Paulev{\'e} and S.~Sen{\'e}.
\newblock {\em {Boolean networks and their dynamics: the impact of updates}}.
\newblock Wiley, Hoboken, 2022.

\bibitem{Perrot2020}
K.~Perrot, M.~Montalva-Medel, P.~P.~B. de~Oliveira, and E.~L.~P. Ruivo.
\newblock {Maximum sensitivity to update schedules of elementary cellular
  automata over periodic configurations}.
\newblock {\em Natural Computing}, 19:51--90, 2020.

\bibitem{Perrot2024a}
K.~Perrot, S.~Sen{\'e}, and L.~Tapin.
\newblock {Combinatorics of block-parallel automata networks}.
\newblock In {\em SOFSEM 2024: Theory and Practive of Computer Science}, volume
  14519 of {\em Lecture Notes in Computer Science}, pages 442--455, Cham, 2024.
  Springer.

\bibitem{Perrot2024b}
K.~Perrot, S.~Sen{\'e}, and L.~Tapin.
\newblock {Complexity of Boolean automata networks under block-parallel update
  modes}.
\newblock In {\em 3rd Symposium on Algorithmic Foundations of Dynamic Networks
  (SAND 2024)}, volume 292 of {\em Leibniz International Proceedings in
  Informatics}, pages 19:1--19:19, Dagstuhl, 2024. Schloss Dagstuhl –
  Leibniz-Zentrum für Informatik.

\bibitem{Perrot2026}
K.~Perrot, S.~Sen{\'e}, and L.~Tapin.
\newblock {Creation of fixed points in block-parallel Boolean automata
  networks}.
\newblock In {\em 22th International Conference on Unconventional Computation
  and Natural Computation (UCNC 2025)}, volume 16364, pages 132--146, Cham,
  2026. Springer.

\bibitem{Rachdi2020}
M.~Rachdi, J.~Waku, H.~Hazgui, and J.~Demongeot.
\newblock {Entropy as a robustness marker in genetic regulatory networks}.
\newblock {\em Entropy}, 22:260, 2020.

\bibitem{Robert1986}
F.~Robert.
\newblock {\em Discrete Iterations: A Metric Study}, volume~6 of {\em Springer
  Series in Computational Mathematics}.
\newblock Springer, 1986.

\bibitem{Ruelle1981}
D.~Ruelle.
\newblock {Small random perturbations of dynamical systems and the definition
  of attractors}.
\newblock {\em Communications in Mathematical Physics}, 82:137--151, 1981.

\bibitem{Ruivo2020}
E.~L.~P. Ruivo, P.~P. Balbi, M.~Montalva-Medel, and K.~Perrot.
\newblock {Maximum sensitivity to update schedules of elementary cellular
  automata over infinite configurations}.
\newblock {\em Information and Computation}, 274:104538, 2020.

\bibitem{Sinai1981}
Y.~G. Sinai.
\newblock {The stochasticity of dynamical systems}.
\newblock {\em Selecta Mathematica Sovietica}, 1:100--119, 1981.

\bibitem{Smale1967}
S.~Smale.
\newblock {Differentiable dynamical systems}.
\newblock {\em Bulletin of the American Mathematical Society}, 73:747--817,
  1967.

\bibitem{Tarjan1972}
R.~E. Tarjan.
\newblock {Depth-first seach and linear graph algorithms}.
\newblock {\em SIAM Journal on Computing}, 1:146--160, 1972.

\bibitem{Thom1972}
R.~Thom.
\newblock {\em Stabilit{\'e} structurelle et morphogen{\`e}se}.
\newblock InterEditions, Paris, 1972.

\bibitem{Thomas1973}
R.~Thomas.
\newblock {Boolean formalization of genetic control circuits}.
\newblock {\em Journal of Theoretical Biology}, 42:563--585, 1973.

\bibitem{Tomassone1993}
R.~Tomassone, C.~Dervin, and J.-P. Masson.
\newblock {\em Biom{\'e}trie\:: Mod{\'e}lisation de ph{\'e}nom{\`e}nes
  biologiques}.
\newblock Masson, Paris, 1993.

\bibitem{Vincent2024}
D.~Vincent, F.~Parsopoulou, L.~Martin, C.~Gaboriaud, J.~Demongeot, G.~Loules,
  S.~Fischer, S.~Cichon, A.~E. Germenis, A.~Ghannam, and C.~Drouet.
\newblock {Hereditary angioedema with normal C1 inhibitor associated with
  carboxypeptidase N deficiency}.
\newblock {\em Journal of Allergy and Clinical Immunology: Global}, 3:100223,
  2024.

\bibitem{Williams1974}
R.~F. Williams.
\newblock {Expanding attractors}.
\newblock {\em Publications Math{\'e}matiques de l'Institut des Hautes
  {\'E}tudes Scientifiques}, 43:169--203, 1974.

\bibitem{Zhang2020}
S.~Zhang, P.~Huang, H.~Dai, Q.~Li, L.~Hu, J.~Peng, S.~Jiang, Y.~Xu, Z.~Wu,
  H.~Nie, Z.~Zhang, W.~Yin, X.~Zhang, and J.~Lu.
\newblock {TIMELESS regulates sphingolipid metabolism and tumor cell growth
  through Sp1/ACER2/S1P axis in ER-positive breast cancer}.
\newblock {\em Cell Death \& Disease}, 11:892, 2020.

\end{thebibliography}

\end{document}